\documentclass[twocolumn,notitlepage,aps,pra,superscriptaddress]{revtex4-1}
\usepackage{amsthm}
\usepackage{amsmath}
\usepackage{amssymb}
\usepackage{mathdots}
\usepackage{graphicx}
\usepackage{mathrsfs}

\makeatletter

\@ifundefined{textcolor}{}
{%
 \definecolor{BLACK}{gray}{0}
 \definecolor{WHITE}{gray}{1}
 \definecolor{RED}{rgb}{1,0,0}
 \definecolor{GREEN}{rgb}{0,1,0}
 \definecolor{BLUE}{rgb}{0,0,1}
 \definecolor{CYAN}{cmyk}{1,0,0,0}
 \definecolor{MAGENTA}{cmyk}{0,1,0,0}
 \definecolor{YELLOW}{cmyk}{0,0,1,0}
}

\usepackage{braket}
\usepackage[backref=true,
bookmarksnumbered=true,
bookmarks=true,
bookmarksopen=true,
colorlinks=true,
citecolor=blue,
linkcolor=blue,
anchorcolor=green,
urlcolor=blue,unicode=false]{hyperref}

\renewcommand{\v}[1]{\ensuremath{\mathbf{#1}}} 
 
\newcommand{\abs}[1]{\left| #1 \right|} 
 
\let\baraccent=\= 
\renewcommand{\=}[1]{\stackrel{#1}{=}} 

\DeclareMathOperator{\Tr}{Tr} 
\DeclareMathOperator{\diag}{diag}

\DeclareMathOperator\Sgn{sgn}
\DeclareMathOperator{\spn}{span}

\makeatother

\begin{document}
\title{Boundary Green functions of topological insulators and superconductors}

\author{Yang Peng}
\affiliation{\mbox{Dahlem Center for Complex Quantum Systems and Fachbereich Physik, Freie Universit{\"a}t Berlin, 14195 Berlin, Germany} }
\author{Yimu Bao}
\affiliation{School of Physics, Peking University, Beijing 100871, China}
\author{Felix von Oppen}
\affiliation{\mbox{Dahlem Center for Complex Quantum Systems and Fachbereich Physik, Freie Universit{\"a}t Berlin, 14195 Berlin, Germany} }

\begin{abstract}
Topological insulators and superconductors are characterized by their gapless boundary modes. In this paper, we develop a recursive approach to the boundary Green function which encodes this nontrivial boundary physics. Our approach describes the various topologically trivial and nontrivial phases as fixed points of a recursion and provides direct access to the phase diagram, the localization properties of the edge modes, as well as topological indices. We illustrate our approach in the context of various familiar models such as the Su-Schrieffer-Heeger model, the Kitaev chain, and a model for a Chern insulator. We also show that the method provides an intuitive approach to understand recently introduced topological phases which exhibit gapless corner states. 
\end{abstract}

\maketitle
\section{Introduction}

There is currently considerable interest in topological phases of matter and topological quantum phase transitions. The best studied and simplest variety of topological phases are the independent-fermion phases \cite{Hasan2010,Qi2011,Bernevig2013book} collected and classified in the periodic table and its extensions
\cite{Schnyder2008,Kitaev2009,Ryu2010,Teo2010,Essin2011,Fulga2012,Chiu2016}. These topological phases can be characterized in terms of bulk topological indices and, by the bulk-boundary correspondence, exhibit gapless modes localized at the sample boundaries.

Here, we present a simple and intuitive recursive approach which provides access to these topological phases and their transitions. Our approach focuses on the boundary Green function \cite{Sancho1985,Umerski1997} which encodes the presence or absence of gapless surface modes in an immediate manner. Focusing on tight-binding models, we imagine that the bulk system is constructed by iteratively adding boundary sites (in one dimension) or boundary layers (in higher dimensions).  We can then exploit that the boundary Green function can be computed recursively and, by virtue of the bulk gap, becomes invariant under this recursion in the thermodynamic limit. Thus, the boundary Green function can be obtained as fixed points of the recursion. 

For a typical model which exhibits topologically trivial and nontrivial phases, the recursion will generate several fixed-point boundary Green functions. The physical boundary Green function and hence the phase diagram can be identified from a stability analysis. The various phases correspond to the stable fixed points of the recursion and the approach to the fixed point contains information about the spatial localization of the boundary mode away from the boundary as well as the bulk gap. This has some similarities with a real space renormalization group approach.  

The fixed-point boundary Green function provides access to various observables of interest such as the tunneling density of states. Most importantly, it immediately contains information on the topological invariants by virtue of the fact that it encodes the full boundary spectrum. We can relate our approach more systematically to the general classifications of free-fermion topological phases by establishing a connection to their classification in terms of scattering matrices \cite{Fulga2012}. This can be done by expressing the reflection matrix describing reflection from a gapped phase in terms of its boundary Green function. In this way, we immediately obtain explicit expressions for the topological invariants in terms of the boundary Green function.  

Our approach can be applied to a wide variety of models. Here, we illustrate the method on a number of popular models. For simple cases such as the Su-Schrieffer-Heeger model \cite{Su1979}, the recursion is conveniently implemented in a direct manner. To make the approach more systematic and to simplify its implementation, we find it helpful to introduce an approach based on the transfer matrix \cite{Pichard1981,Lee1981,Hatsugai1993a,Hatsugai1993b}. This is illustrated in one dimension for the Kitaev chain \cite{Kitaev2001}  and in two dimensions for a model of a Chern insulator \cite{Bernevig2013book}. 

Very recently, it was pointed out \cite{Benalcazar2016} that some 2D and 3D models exhibit topologically protected excitations at corners. These models have boundary Green functions which are regular, reflecting the fact that the boundary is gapped. Nevertheless, it is natural to interpret the inverse of the boundary Green function at the Fermi energy as an effective boundary Hamiltonian which can by itself be topologically trivial or nontrivial. We show for a model discussed in Ref.\ \cite{Benalcazar2016} that this effective boundary Hamiltonian is indeed topologically nontrivial when the model exhibits topological corner states. 
Our boundary Green function method may thus open the path to a more systematic analysis of such higher-order topological phases.

The paper is organized as follows. In Sec.~\ref{sec:basic}, we introduce the principal idea of this work and illustrate it with the Su-Schrieffer-Heeger model.  We also show how to relate the boundary Green function to the reflection matrix which connects our approach with the systematic classification of free-fermion systems.  
In Sec.~\ref{sec:GF_TM}, we show how the transfer matrix can be employed for computing the fixed-point boundary Green function. This is used in Sec.~\ref{sec:homogeneous} to compute the boundary Green function for the Kitaev chain and a model of a Chern insulator. For both models, we show how the topological invariants can be extracted from the fixed-point boundary Green function. A 2D model with topological corner states is investigated in Sec.~\ref{sec:inhomogeneous}. We conclude in Sec.~\ref{sec:conclusion}. Some aspects are relegated to appendices.

\section{Basic considerations\label{sec:basic}}

\subsection{Boundary Green function}

\begin{figure}[t!]
    \centering
    \includegraphics[width=0.45\textwidth]{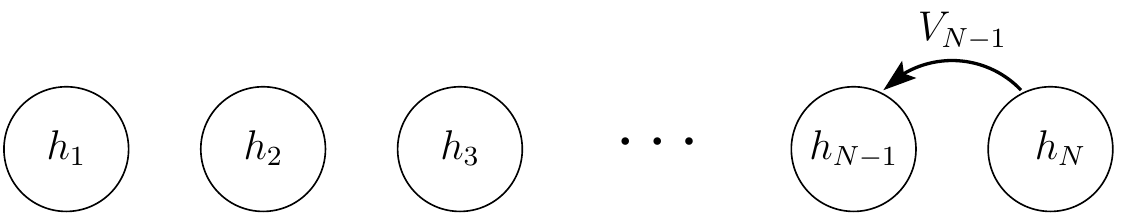}
    \caption{Schematic representation of the quasi-one-dimensional Hamiltonian with $N$ unit cells.}
    \label{fig:tight-binding}
\end{figure}

Consider a $d$-dimensional tight-binding Hamiltonian, which is translationally invariant along $d-1$ directions. Assuming periodic boundary conditions in these directions, we can define a $(d-1)$-dimensional crystal momentum $\v{k}_\perp$ to label the Hamiltonian. In the remaining dimension, we apply open boundary conditions and retain a real-space representation. By choosing a sufficiently large unit cell, a Hamiltonian with a finite hopping range can always be brought into a form which couples only neighboring unit cells. Thus, without loss of generality, we can consider a family of quasi-1D Hamiltonians which are labelled by the transverse momentum $\v{k}_{\perp}$  and take the form
\begin{equation}
    H(\v{k}_\perp) = \sum_{n=1}^{N}\left(\psi_{n}^{\dagger}h_{n}\psi_{n}+\psi_{n+1}^{\dagger}V_{n}\psi_{n}+\psi_{n}^{\dagger}V_{n}^{\dagger}\psi_{n+1}\right).
    \label{eq:modelham}
\end{equation}
Here, $n=1,2\ldots N$ labels the unit cells along the direction with open boundary conditions, and $\psi_{n}$ ($\psi^{\dagger}_n$) is a column (row) vector of electron annihilation (creation) operators in the $n$th unit cell. These vectors have $M$ entries, reflecting the dimension of the local Hilbert space associated with each unit cell and a given $\v{k}_{\perp}$. The intra- and inter-unit-cell couplings are described by $M\times M$ matrices $h_{n}$ and $V_n$, respectively. Note that we suppressed the $\v{k}_\perp$ dependence of $\psi_n$, $h_n$, and $V_n$ for notational simplicity. The structure of this quasi-1D tight-binding Hamiltonan is illustrated in Fig.~\ref{fig:tight-binding}. 

More explicitly, the corresponding first-quantized Hamiltonian can be written as an $NM\times NM$ block tridiagonal Hamiltonian matrix $\v{H}(\v{k}_{\perp})$ which takes the form
\begin{equation}
    \mathbf{H}=\left(\begin{array}{ccccc}
        h_{1} & V_{1}^{\dagger}\\
        V_{1} & h_{2} & V_{2}^{\dagger}\\
         & \ddots & \ddots & \ddots\\
          &  & V_{N-2} & h_{N-1} & V_{N-1}^{\dagger}\\
           &  &  & V_{N-1} & h_{N}
       \end{array}\right).
       \label{Htridiag}
\end{equation}
Both the spectrum and the eigenfunctions of the system are encoded in the Green function (or resolvent operator) $\mathbf{G}(\omega)$ of this Hamiltonian, which is defined through
\begin{equation}
    (\omega \mathbb{I} - \v{H})\mathbf{G}(\omega) = \mathbb{I}.
    \label{eq:def_G}
\end{equation}
Here, $\mathbb{I}$ is the identity matrix  of dimension $NM$. 

The low-energy excitations of topological insulators and superconductors are localized near their boundary. This motivates us to consider the boundary Green functions $G_{1}= \mathbf{G}_{11}$ and $G_{N}=\mathbf{G}_{NN}$, which correspond to the $1,1$ and $N,N$ blocks of the full Green function matrix $\mathbf{G}$. Thus, these boundary Green functions are $M\times M$ matrices.

By virtue of the tridiagonal structure of the first-quantized Hamiltonian (\ref{Htridiag}), we can compute the boundary Green function recursively, extending the system unit cell by unit cell in the direction with open boundary conditions. In this way, we can relate the boundary Green function $G_N$ of a chain with $N$ unit cells to the boundary Green function $G_{N-1}$ of a chain with $N-1$ unit cells. Indeed, the two boundary Green functions are simply related by the Dyson equation
\begin{equation}
    \left(g_{N}^{-1} - V_{N-1}G_{N-1}V_{N-1}^{\dagger} \right) G_{N} = \mathbb{I}.
    \label{eq:Dyson}
\end{equation}
Here, $g_{N}^{-1}(\omega) = \mathbb{I}\omega- h_{N}$ is the bare Green function of the $N$th unit cell. 
Similar recursions were used in the context of topological phases in Refs.~\cite{Dai2008,Kim2015}.

When considering insulators and superconductors with a gapped bulk, be they topological or not, we expect that the boundary Green function becomes independent of the number of layers in the limit of large $N$. Hence, the boundary Green function should approach a fixed point $G$ which satisfies the fixed-point equation
\begin{equation}
    \left(g^{-1} - VGV^{\dagger} \right) G = \mathbb{I}.
    \label{eq:fxpt}
\end{equation}
Here, we have assumed that the system is uniform along the chain, so that we can drop the indices of $g$ and $V$. This closed equation for the boundary Green function (a matrix quadratic equation) is the central starting point of our considerations. 

\subsection{The Su-Schrieffer-Heeger (SSH) model: An illustrative example}

It should be evident that the boundary Green function encodes much information of interest in the context of topological insulators and superconductors. Most importantly, it knows about gapless end, edge, or surface states, and can thus be used to derive their properties (including their extension into the bulk) as well as the  
topological phase diagram of the model. It is useful to illustrate this statement in the context of the Su-Schrieffer-Heeger model. This model is well known to exhibit topologically distinct phases and is sufficiently simple so that we can implement the recursive approach in a rather straight-forward manner, without relying on the more powerful methods developed in the next sections. 

The Su-Schrieffer-Heeger (SSH) model describes a 1D chain of spinless fermions with one orbital per site and
alternating hopping strengths between neighbouring sites. For a chain with $2N$ sites and open boundary conditions, the Hamiltonian takes the form
\begin{equation}
    H = \sum_{n}\left( -t_1c_{2n}^{\dagger}c_{2n-1} - t_2 c_{2n+1}^{\dagger} c_{2n} + {\rm h.c.}\right),
    \label{eq:SSH}
\end{equation}
where $c_{j}$ ($c_{j}^{\dagger}$) annihilates (creates) a fermion at site $j$, and $t_1$ and $t_2$ are the hopping amplitudes (taken to be real for simplicity). While the model is gapless for $\abs{t_1}=\abs{t_2}$, the dimerization opens a gap around $E=0$ when the hopping amplitudes differ in magnitude. The gapless point separates a topological phase for $\abs{t_{1}/t_{2}}<1$ from a trivial phase for $\abs{t_{1}/t_{2}}>1$. In the topological phase, there is a midgap state localized at each end of the chain in the thermodynamic limit.

For the SSH  model, we have $g_N^{-1} = \omega$ and 
\begin{gather}
V_N = \begin{cases}
t_1 & \quad {\rm odd} \ N \\
t_2 & \quad {\rm even} \ N
\end{cases}
\end{gather}
from Eq.\ (\ref{eq:SSH}). In view of the dimerization of the hopping amplitudes, it is convenient to  
iterate the recursion Eq.\ (\ref{eq:Dyson}) and to obtain a recursion for chains with an even number of sites, 
\begin{equation}
G_{2N} = [\omega-t_1^2(\omega-t_2^2G_{2N-2})^{-1}]^{-1}.
\label{eq:SSH_rec}
\end{equation}
To analyze this recursion, we write it as
\begin{equation}
G_{2N} - G_{2N-2} = \beta(G_{2N-2})
\end{equation}
with the $\beta$-function
\begin{equation}
\beta(x) = [\omega-t_1^2(\omega-t_2^2x)^{-1}]^{-1}-x.
\label{eq:SSH_beta}
\end{equation}
The fixed-point boundary Green function $G$ follows from the zeroes of the $\beta$-function. Solving the corresponding quadratic equation, we  find the  two solutions
\begin{eqnarray}
    G_{\rm triv} &=& \frac{\omega}{t_2^2-t_1^2} + O(\omega^3), \label{eq:G_SSH_triv} \\
G_{\rm top} &=&\frac{t_2^2-t_1^2}{t_2^2\omega} + O(\omega) , 
\label{eq:G_SSH_top}
\end{eqnarray}
where we have restricted attention to the limit of small $\omega$. 

These two fixed-point boundary Green functions correspond to the topological and nontopological phases of the model. In the topological phase, the midgap state makes the boundary Green function singular for $\omega\to0$. In contrast, there is no such singularity in the trivial phase where the end remains gapped and the boundary Green function is regular. These statements are an immediate consequence of the Lehmann representation for the Green function.

The phase diagram of the model emerges when studying the stability of the fixed-point Green functions under the recursion (\ref{eq:SSH_rec}). A fixed point is stable when $\beta^\prime(x) < 0$. A simple calculation yields 
\begin{eqnarray}
  \beta^\prime(G_{\rm top}) &=& \frac{t_1^2}{t_2^2} -1 \\
  \beta^\prime(G_{\rm triv}) &=& \frac{t_2^2}{t_1^2} -1
\end{eqnarray}
for $\omega=0$. This implies that the singular boundary Green function $G_{\rm top}$ corresponding to the topological phase is stable for $\abs{t_{1}/t_{2}}<1$, while the regular boundary Green function $G_{\rm triv}$ corresponding to the trivial phase is stable for $\abs{t_{1}/t_{2}}>1$.

The fixed-point analysis also provides information about the gap, the correlation length of the model, and their behavior near the topological critical point  $t_1=t_2$. The correlation length of the phase with fixed-point Green function $G$ is given by $\xi=4/|\beta^\prime(G)|$. (Notice that we define $\xi$ as the decay length of a wavefunction; the diagonal elements of the Green function decay on half of this scale.) Thus, we find the result
\begin{equation}
  \xi = \frac{4[{\rm max}(t_1,t_2)]^2}{|t_1^2-t_2^2|},
\end{equation}
which is written in a way that applies to both phases. The correlation length diverges at the topological critical point $|t_1|=|t_2|$. It is also  interesting to note that this result is consistent with the quasiparticle weight $Z$ of the fixed-point Green function. In the topological phase, one expects $G_{\rm top} \simeq Z/\omega$ for small $\omega$ and $Z = 4/\xi$ reflecting the fact that the midgap state is rapidly oscillating and decays exponentially into the bulk on the scale of the correlation length. 

To extract the spectral gap $\Delta$, we note that it is related to the correlation length through $\Delta=\hbar v_F/\xi$ (assuming that the gap is small compared to the bandwidth, i.e., $|t_1-t_2| \ll t_1,t_2$). Here, the Fermi wavelength should be taken at the center of the band with $t=t_1=t_2$, i.e., $v_F=2t/\hbar$. This yields $\Delta = |t_1 - t_2|$ in agreement with a direct evaluation of the spectrum. 

\subsection{Topologial invariants}

The boundary Green function contains direct information on the topological nature of the phases through the  gapless end, edge, or surface states described by the boundary Green function. As for the SSH model above, this can be read off directly from the boundary Green function at energies $\omega$ below the bulk gap. In 1D, the topological phase is signaled by midgap end states and hence by a contribution to the Green function which is singular for $\omega\to 0$. In higher dimensions, there are edge or surface modes with a linear dispersion at low energies. We will discuss this in more detail below. 

A systematic connection to the periodic table of topological phases can be established by relating the boundary Green function to the reflection matrix. Ref.~\cite{Fulga2012} derived the periodic table based on the reflection matrix $r_N$ and provided explicit expressions for the topological invariants in terms
of the $r_N$. With the relation between the reflection matrix and the boundary Green function, the topological invariants can alternatively be written in terms of the latter. 

Start with a tight-binding model of the kind discussed above and attach a lead to one end with $M$ propagating channels for each ${\mathbf k}_\perp$. In the spirit of the tight-binding model, the channels are only coupled to the last site of the chain, with the matrix $W$ denoting the coupling between the lead channels and the adjacent site of the system. By the gapped nature of the system, the incoming wave is fully reflected. Based on the Mahaux-Weidenm\"uller formula, the corresponding scattering -- or reflection -- matrix $r_N$ can then be written as \cite{Mahaux1969}
\begin{equation}
   r_N = \mathbb{I} - 2i W \frac{1}{G_{N}^{-1}+iW^\dagger W}W^\dagger.
\end{equation}
Note that the Green function $G_N$ appearing in this expression is indeed the boundary Green function since the coupling is to the last site of the chain only. Some rearranging then leads to the relation \cite{Aleiner2002}
\begin{equation}
r_{N}=\frac{\mathbb{I}-iWG_{N}W^{\dagger}}{\mathbb{I}+iWG_{N}W^{\dagger}}.
\label{eq:rG}
\end{equation}
We include an alternative and more detailed derivation of this relation in App.\ \ref{app:BGF_R}. Moreover, we couch the scattering-matrix approach to the periodic table and the topological invariants in terms of the boundary Green function in App.\ \ref{app:table}.

Before concluding this section, we briefly consider systems with a trivial bulk topological invariant and thus without a gapless boundary mode. In this case, the boundary Green function evaluated at $\omega=0$ is  generally invertible. This can be used to define an effective boundary Hamiltonian 
\begin{equation}
    H_{\rm bound} = -\left[G(\omega=0)\right]^{-1}.
\end{equation}
Even if the original model does not have a nontrivial bulk topological invariant, this boundary Hamiltonian can still be topologically nontrivial, leading to a second order topological phase. We use this observation to extend our approach to describe higher-order topological phases in Sec.~\ref{sec:2nd_order}.

\section{Boundary Green function and transfer matrix\label{sec:GF_TM}}

In this  section, we introduce a more efficient method to compute the fixed-point boundary Green function which comes in handy when discussing models which are more complicated
than the SSH model. The method is based on the transfer matrix \cite{Pichard1981,Lee1981} and effectively provides an explicit construction of the fixed-point boundary Green function. 

We start by reviewing the transfer matrix based on the Schr\"odinger equation for the Hamiltonian in Eq.\ (\ref{eq:modelham}),
\begin{equation}
    g_{n}^{-1}(\omega)\psi(n)-V_{n-1}\psi(n-1)-V_{n}^{\dagger}\psi(n+1)=0.
    \label{eq:SEQ}
\end{equation}
Here, ${\Psi}(n)$ is an $M$-component wave function. The site index $n$ takes on values $n=1,\dots,N$, and we impose the open boundary conditions $\psi_j(0)=0=\psi_j(N+1)$. 

As usual for tight-binding Hamiltonians, we can relate the two-component quantities $\Psi(n) =[\psi(n+1)^T,\psi(n)^T]^T$ via $\Psi(n)=M_{n}(\omega)\Psi(n-1)$, where the matrix $M_{n}(\omega)$ takes the form 
\begin{equation}
    M_{n}(\omega)=\left(\begin{array}{cc}
        (V_{n}^{\dagger})^{-1}g_{n}^{-1}(\omega) & -(V_{n}^{\dagger})^{-1}V_{n-1}\\
        \mathbb{I} & 0
    \end{array}\right)
    \label{eq:Mj}.
\end{equation}
Here and in the following, we assume $V_n$ to be invertible. Notice that $V_{0}$ and $V_{N}$ do not appear in the Hamiltonian and can be chosen as any invertible matrix in order to define $M_{0}$ and $M_{N}$. The matrix $M_{n}(\omega)$ is sometimes called a transfer matrix. Here, we reserve this terminology for the matrix   
\begin{equation}
    \mathcal{M}_{N} =  M_N M_{N-1} \dots M_1,
\end{equation}
connecting the two ends of the chain. 

If $\omega_0$ is an eigenenergy of the Schr\"odinger equation (\ref{eq:SEQ}), the transfer matrix $\mathcal{M}_{N}(\omega_0)$ connects the eigenfunctions $\psi(n)$ at sites $n=1$ and $n=N$,
\begin{equation}
    \left(\begin{array}{c}
        0\\
        \psi(N)
    \end{array}\right)=\mathcal{M}_{N}(\omega_{0})\left(\begin{array}{c}
        \psi(1)\\
        0
    \end{array}\right).
\end{equation}
If the state corresponding to the eigenenergy is $k$-fold degenerate, ${\Psi}_{p}(n)$ with $p=1,\dots,k$, then 
\begin{equation}
    \det\mathcal{M}_{N,11}(\omega_0)  = 0,
\end{equation}
and 
\begin{equation}
    \ker \mathcal{M}_{N,11}(\omega_0) = \spn \{\psi_1(1),\dots, \psi_k(1)\}.
\end{equation}
Here, the subscript $11$ denotes the blocks under the partition of $\mathcal{M}_N$ into a $2\times2$ block matrix. One immediate consequence of this is that $k \leq M$, namely the degeneracy is bounded by the number of internal degrees of freedom per unit cell.

We now show how the boundary Green function $G_N$ can be expressed in terms of the transfer matrix $\mathcal{M}_{N}$, starting directly from the recursion in Eq.~(\ref{eq:Dyson}). Defining $X_n = G_n V^\dagger_n$, we can rewrite the recursion as
\begin{equation}
  (V_{N}^{\dagger})^{-1}g_N^{-1} X_N - (V^{\dagger}_N)^{-1}V_{N-1}X_{N-1}X_{N} = \mathbb{I},
\end{equation}
or equivalently
\begin{equation}
\left(\begin{array}{c}
  \mathbb{I}\\
  X_{N}
\end{array}\right)=M_{N}\left(\begin{array}{c}
  \mathbb{I}\\
  X_{N-1}
\end{array}\right)X_{N},
\end{equation} 
where $M_N$ is defined in Eq.~(\ref{eq:Mj}). 

This relation can be applied iteratively to obtain
\begin{align}
 \left(\begin{array}{c}
  \mathbb{I}\\
  X_{N}
\end{array}\right)= \mathcal{M}_{N}\left(\begin{array}{c}
  \mathbb{I}\\
  0
\end{array}\right)X_1\dots X_N.
\end{align}
The first row of this equation implies that
\begin{equation}
  X_{1}\dots X_N = \mathcal{M}_{N,11}^{-1}.
\end{equation}
Then, the second row yields
\begin{align}
  X_N = \mathcal{M}_{N,21} X_{1}\dots X_N = \mathcal{M}_{N,21}\mathcal{M}_{N,11}^{-1}.
\end{align}
Hence, we obtain the desired expression 
\begin{equation}
  G_N =  \mathcal{M}_{N,21}\mathcal{M}_{N,11}^{-1} (V_N^\dagger)^{-1}
 \label{eq:GNbound}
\end{equation}
for the boundary Green function in terms of the transfer matrix. This formula was also derived in the broader context of block Toeplitz matrices in Ref.\ \cite{Boffi2014}.

It is useful to exploit the symplectic structure of  the transfer matrix. Notice that the matrix $M_n$ has the property
\begin{equation}
    M_{n}^\dagger \Omega_{n} M_{n} = \Omega_{n-1}
\end{equation}
in terms of the anti-hermitian matrix
\begin{equation}
    \Omega_{n}=\left(\begin{array}{cc}
        0 & -V_{n}\\
        V_{n}^{\dagger} & 0
    \end{array}\right).
\end{equation}
As a consequence, the transfer matrix $\mathcal{M}_N$ obeys the relation
\begin{equation}
    \mathcal{M}_{N}^{\dagger}\Omega_{N}\mathcal{M}_N = \Omega_{0}.
\end{equation}
Remember that the choice of matrices $V_N$ and $V_0$ at the boundary is arbitrary as long as they are invertible. If we choose $V_0=V_N$, then the transfer matrix obtains a symplectic structure as it satisfies the relation
\begin{equation}
    \mathcal{M}_{N}^{\dagger}\Omega_{0}\mathcal{M}_N = \Omega_{0}.
\end{equation}
As usual for symplectic matrices, this implies that the eigenvalues of $\mathcal{M}_{N}$ appear in pairs of $\lambda$ and $1/\lambda^{*}$. In particular, if $v$ is a right eigenvector of $\mathcal{M}_{N}$ with eigenvalue $\lambda$,
\begin{equation}
    \mathcal{M}_N v = \lambda v,
\end{equation}
then $(\Omega_0 v)^\dagger$ is a left eigenvector of $\mathcal{M}_{N}$ with eigenvalue $1/\lambda^*$,
\begin{equation}
    (\Omega_0 v)^\dagger \mathcal{M}_N  = \frac{1}{\lambda^*} (\Omega_0 v)^\dagger.
\end{equation}
This separates the $2M$ eigenvalues of the transfer matrix into $M$ eigenvalues $\lambda_j$ ($j=1,\ldots, M$) with $|\lambda_j|>1$, and another $M$ eigenvalues $1/\lambda_j^*$ with modulus smaller than unity. The Lyapunov exponents $1/\xi_j$ are then defined through 
\begin{equation}
    \frac{1}{\xi_{j}}  = \lim_{N\to\infty}\frac{1}{N}{\ln \abs{\lambda_{j}}}
\end{equation}
and are strictly positive at the Fermi energy ($\omega=0$) for a gapped bulk \cite{Beenakker1997}. At the topological phase transition, the bulk gap closes and therefore at least one Lyapunov exponent vanishes for some $j$, corresponding to $\abs{\lambda_j}=1$ or a diverging $\xi_{j}$. This is explicitly shown in App.~\ref{app:Lyapunov}.

So far, we did not assume that the system is uniform along the direction with open boundary conditions. Let us now consider models with translational invariance of period $a$. (In the SSH model discussed above, this period was equal to 2 due to the dimerization of the hopping.) Considering a chain of length $Na$, the transfer matrix becomes 
\begin{equation}
    \mathcal{M}_{aN} = T^{N}, 
    \label{eq:MT}
\end{equation}
where we defined the matrix
\begin{equation}
    T = M_{a} M_{a-1}\dots M_{1}
\end{equation}
associated with one period of length $a$. 

In the translationally invariant case, the evaluation of the boundary Green function can be simplified by diagonalizing the matrix $T$. Again, $T$ satisfies a symplectic structure and its eigenvalues can be separated into $\Lambda = \diag(\lambda_1,\dots,\lambda_M)$ with $\abs{\lambda_j}>1$, and a second set $(\Lambda^{-1})^*$. Thus, we can write
\begin{equation}
    T\left(\begin{array}{cc}
        U_{11} & U_{12}\\
        U_{21} & U_{22}
    \end{array}\right)=\left(\begin{array}{cc}
        U_{11} & U_{12}\\
        U_{21} & U_{22}
    \end{array}\right)\left(\begin{array}{cc}
        \Lambda & 0\\
        0 & (\Lambda^{-1})^*
    \end{array}\right)
    \label{eq:eigenT}
\end{equation}
We can use this to express $T$ and hence $\mathcal{M}_{aN}$, and insert this expression into the Eq.\ (\ref{eq:GNbound}) for the boundary Green function. Upon taking the thermodynamic limit
$N\to \infty$ and using that the system is gapped (see App.~\ref{app:proveM_U} for details), this yields the desired explicit expression 
\begin{equation}
  G = \lim_{N\to\infty}G_{aN} =   U_{21}U_{11}^{-1} (V_{0}^{\dagger})^{-1} ,
    \label{eq:Ninf}
\end{equation}
for the fixed-point boundary Green function (see also Ref.~\cite{Boffi2014}).  While we have suppressed the argument $\omega$ in this expression, it is valid for any $\omega$ which falls inside the bulk gap. 

For a homogeneous system in which the on-site potential $h_n$ and the hopping amplitude $V_n$ are independent of the site index $n$ along the chain, i.e., a system with $a=1$, the evaluation of the boundary Green functions can be simplified further. In this case, the columns of the matrix $U$ in Eq.\ (\ref{eq:Ninf}) consist of the right eigenvectors of 
\begin{equation}
 T = \left(\begin{array}{cc}
        (V^{\dagger})^{-1}g^{-1} & -(V^{\dagger})^{-1}V\\
        \mathbb{I} & 0
    \end{array}\right)
    \label{eq:M_homogeneous}.
\end{equation}
The simple structure of $T$ implies that the right eigenvector with eigenvalue $\lambda_i$ can be written as $(\lambda_i z_{i}^T,z_{i}^T)^T$, where  the vector $z_i$ obeys
\begin{equation}
  \left( \frac{V}{\lambda}_i + V^\dagger \lambda_i - g^{-1} \right) z_i = 0.
  \label{eq:eqforx}
\end{equation}
Let us also denote the eigenvectors corresponding to eigenvalues $|\lambda_i|>1$ as $x_i$ and the corresponding eigenvectors with eigenvalues $1/\lambda_i^*$ as $y_i$. We can now identify $U_{21}=(x_1,\ldots,x_M)$ and $U_{11}=(\lambda_1 x_1,\ldots, \lambda_M x_M)$. Then, Eq.\ (\ref{eq:Ninf}) implies 
\begin{equation}
  G =  \left(x_{1}, \dots, x_{M}\right)\Lambda^{-1}(x_1,\dots,x_M)^{-1} (V^\dagger)^{-1}.
  \label{eq:G_homogeneous}
\end{equation}
for the fixed-point boundary Green function. The essential steps in evaluating $G$ thus reduce to solving the nonlinear eigenvalue problem Eq.\ (\ref{eq:eqforx}) and inserting the result into Eq.\ (\ref{eq:G_homogeneous}). 

For later use, it is useful to collect expressions for all right and left eigenvectors of $T$ for $a=1$. We write the right eigenvectors of $T$ as $v_i=(x_i^T,x_i^T/\lambda_i)^T$ for $|\lambda_i|>1$ and as $v_{M+i} =
(y_i^T,y_i^T\lambda_i^*)^T$ for
the corresponding eigenvalues $1/\lambda_i^*$ with modulus less than unity. The symplectic nature of $T$ implies then
that the corresponding left eigenvectors  take the form $w_i^T=(-y_{i}^{\dagger}V^\dagger\lambda_i,y_{i}^{\dagger}V)$
with eigenvalue $\lambda_i$ and  $w^T_{M+i}=(-x_{i}^{\dagger}V^\dagger/\lambda_i^*,x_{i}^{\dagger}V)$ with eigenvalue $1/\lambda_i^*$.

\section{Application to 1D and 2D models \label{sec:homogeneous}}

In this section, we apply the formalism developed in the previous section to two homogeneous tight-binding models with $a=1$. We consider the Kitaev chain as an example of a 1D model and a Chern insulator as an example of a 2D model. 

\subsection{Perturbation theory}
\label{sec:pert}

We are mainly interested in the fixed-point boundary Green function for small $\omega$ as this limit captures the topological properties and the effective low-energy excitations. This limit can be most directly treated by means of perturbation theory. Denoting $T_{0}=T(\omega=0)$, we have
\begin{equation}
    T=T_{0}+\omega A,\quad A=\left(\begin{array}{cc}
      (V^{\dagger})^{-1} & 0\\
        0 & 0
    \end{array}\right).
\end{equation}
For small $\omega$, we can treat $\omega A$ in first-order perturbation theory (properly generalized to nonhermitian matrices, for which left and right eigenvectors are no longer identical).

If we denote by $w_{i}^{T}$ and $v_{i}$ the left and right eigenvectors of $M_{0}$ with eigenvalues  $\lambda_i$ (not restricted in magnitude), we obtain the first-order corrections
\begin{equation}
    \delta\lambda_{i}=\omega\frac{w_{i}^{T}Av_{i}}{w_{i}^{T}v_{i}},\quad\delta v_{i}=\omega\sum_{j\neq
    i}\frac{v_{j}}{\lambda_{i}-\lambda_{j}}\frac{w_{j}^{T}Av_{i}}{w_{j}^{T}v_{j}}.
\end{equation}
These expressions are evident generalizations of the corresponding results in quantum mechanics. 

We can now use the explicit structure of the $v_i$ discussed at the end of the previous section as well as the simple form of $A$. This yields 
\begin{equation}
    \delta\lambda_{i}=-\frac{\omega\lambda_{i}y_{i}^{\dagger}x_{i}}{y_{i}^{\dagger}(V/\lambda_{i}-V^\dagger\lambda_{i})x_{i}}
\end{equation}
for the shift in the eigenvalues with $\abs{\lambda_{i}}>1$ and 
\begin{equation}
    \delta v_{i}=\omega\left(\begin{array}{c}
        \delta r_{i}\\
        \delta s_{i}
    \end{array}\right),
\end{equation}
with 
\begin{align}
    \delta r_{i}&=-\sum_{j\neq
    i}\frac{y_{j}^{\dagger}x_{i}}{(\lambda_{i}\lambda_j^*-1)y_{j}^{\dagger}(V/\lambda_{j}-V^\dagger\lambda_{j})x_{j}}x_{j} \nonumber \\
    &-\sum_{j}\frac{x_{j}^{\dagger}x_{i}}{(\lambda_{i}\lambda_j^*-1)x_{j}^{\dagger}(V\lambda_{j}^{*}-V^\dagger/\lambda_{j}^{*})y_{j}}y_{j},
\end{align}
for the corresponding first-order corrections to the right eigenvectors. In the last expression, the sum is only over those indices $j$ with $\abs{\lambda_{j}}>1$.

At small $\omega$, the fixed-point boundary Green function in Eq.~(\ref{eq:G_homogeneous}) then becomes
\begin{align}
  G &=(x_{1},\dots,x_{M}) \Lambda^{-1}\nonumber \\
    &\times \left(x_{1}+\omega \delta r_{1},\dots,x_{M}+\omega\delta r_{M}\right)^{-1} (V^\dagger)^{-1}.
\end{align}
Notice that we retain the correction due to a nonzero $\omega$ only in the matrix inverse because this is where a singular $\omega$ dependence can originate from. Notice also that $\delta s_i$ does not enter this expression so that we refrain from giving an explicit expression. 

\subsection{Kitaev chain}

\subsubsection{Model and boundary Green function}

As a first example, consider the Kitaev chain which is a lattice model for a spinless $p$-wave superconductor. 
The Hamiltonian for a chain of $N$ sites takes the form 
\begin{equation}
    H = -\mu \sum_{j=1}^{N}c_{j}^{\dagger}c_{j} + \sum_{j=1}^{N-1}\left[t c_{j+1}^{\dagger}c_{j} + \Delta c_{j+1}^{\dagger}c_j^{\dagger} + {\rm h.c.}\right], 
    \label{eq:H_Kitaev}
\end{equation}
where $c_{j}$ ($c_j^{\dagger}$) is the fermion annihilation (creation) operator at site $j$. We assume below that both the hopping $t$ and the pairing $\Delta$ are real.

This model has trivial and topological phases, with the latter hosting Majorana zero modes at both ends of the chain. The trivial phase occurs for $\abs{\mu}>2\abs{t}$ when the chemical potential $\mu$ falls outside the normal-state band. The topological phase occurs for $\abs{\mu}<2\abs{t}$. The model supports a chiral symmetry $\mathcal{U}_S = \tau_x$, so that it falls into class BDI. Indeed, in addition to the topological phase transition lines at $\abs{\mu}=2\abs{t}$, the model also becomes gapless in the absence of pairing correlations, $\Delta=0$, and the topological phases with $\Delta>0$ and $\Delta<0$ have topological indices $\pm 1$.

To apply the approach developed in Sec.~\ref{sec:GF_TM}, we introduce a Nambu spinor $\psi_{j}^{\dagger}=(c_{j}^{\dagger},c_{j})$. Then, we can identify 
\begin{gather}
    g^{-1}(\omega) = \omega + \mu\tau_{z} \\
    V = t\tau_z + i \Delta\tau_y,
\end{gather}
where $\tau_{x,y,z}$ are Pauli matrices in Nambu space. 

Applying the perturbative approach developed in Sec.~\ref{sec:pert}, we first determine the eigenvalues and eigenvectors of $T$ for $\omega = 0$. The eigenvalues follow from the condition
\begin{equation}
    \det\left[ V/\lambda+V^{\dagger}\lambda-g^{-1}(\omega=0)\right]= 0,
\end{equation}
which yields
\begin{equation}
    (t\mp \Delta)\lambda^2 - \mu \lambda + (t\pm \Delta) =0.
\end{equation}
Denote the pair of roots for the upper sign as $\lambda_{1,2}$, the roots for the lower sign as $\lambda_{3,4}$, and fix the labeling by imposing $\lambda_{1}\lambda_{4}^{*}=1=\lambda_{2}\lambda_{3}^{*}$.
The corresponding eigenvectors fulfilling Eq.~(\ref{eq:eqforx}) are
\begin{equation}
    z_{1}=z_{2}=\left(\begin{array}{c}
        1\\
        1
    \end{array}\right),
    \quad
    z_{3}=z_{4}=\left(\begin{array}{c}
        1\\
        -1
    \end{array}\right).
\end{equation}
To determine the two eigenvectors corresponding to eigenvalues with $|\lambda|>1$, we consider the quantity 
\begin{align}
 (\lambda_1^2 -1)(\lambda_2^2 -1) 
=(\lambda_1\lambda_2+1)^2 - (\lambda_1 + \lambda_2)^2 
= \frac{4t^2 - \mu^2}{(t-\Delta)^2}
\end{align}
Thus, $\abs{\lambda_1}$ and $\abs{\lambda_2}$ are both larger than or both smaller than unity when $\abs{2t}>\abs{\mu}$.

Let us first consider the situation that $\abs{\lambda_1}$ and $\abs{\lambda_2}$ are both larger than unity. This occurs for $\abs{2t}>\abs{\mu}$ and $\Delta t<0$. Then, the matrix $(x_1,x_2)=1+\tau_x$ is noninvertible and the fixed-point boundary Green function $G$ becomes singular in the limit $\omega\to0$. This is in agreement with expectations as the model exhibits a topological phase for these parameters. 

Including the first-order corrections in $\omega$, we find 
\begin{equation}
    G(\omega) =  \frac{(1+\tau_{x})}{2\omega(t-\Delta)a_{12}}
    \label{eq:fpgfkitaevtop}
\end{equation}
where
\begin{align}
    a_{12}  &=  \frac{1}{(\abs{\lambda_1}^2-1)(\lambda_2 - \lambda_4)(\lambda_1^*(t-\Delta) - \lambda_4(t+\Delta))}  \nonumber \\
    & + \frac{1}{(\abs{\lambda_2}^2-1)(\lambda_1 - \lambda_3)(\lambda_2^*(t-\Delta) - \lambda_3(t+\Delta))}. 
\end{align}
Thus, the fixed-point boundary Green function takes a form which implies the existence of a zero-energy end state. The rank of the numerator $1+\tau_x$ confirms that there is one zero-energy end state at each end, and the matrix structure imposes that the end-state wavefunction has particle-hole symmetry as it should for a Majorana bound state. The boundary Green function of the Kitaev chain was also derived in Ref.~\cite{Zazunov2016} by a different method.

As for the SSH model, the quasiparticle weight of the fixed-point boundary Green function (\ref{eq:fpgfkitaevtop}) encodes information on the localization length of the Majorana end state. In particular, one expects that the Majorana localization length diverges at the topological phase transition. For the parameters of Eq.\ (\ref{eq:fpgfkitaevtop}), we have either $\abs{\lambda_{1}}\to 1$ or $\abs{\lambda_2} \to 1$, so that $a_{12}$ is dominated by one of the two terms and does indeed diverge. 

Similarly, we find $\abs{\lambda_{3,4}}>1$ for $\abs{2t}>\abs{\mu}$ and $\Delta t>0$. Then, the matrix $(x_1,x_2)=1-\tau_x$ (exploiting that the eigenvectors are only defined up to an overall prefactor) and we find 
\begin{equation}
    G(\omega) =  \frac{(1-\tau_{x})}{2\omega(t+\Delta)a_{34}}
    \label{eq:G34}
\end{equation}
where $a_{34}$ follows from the expression for $a_{12}$ by the replacements $(1,2) \leftrightarrow (3,4)$. 
As expected, this expression is again consistent with the particle-hole symmetry of Majorana end states. 

Now, consider the nontopological phase in the complementary parameter range $\abs{2t}<\abs{\mu}$. Assuming for definiteness that $\abs{\lambda_{1,3}}>1$, we have the relation 
\begin{equation}
    \frac{\lambda_1}{t+\Delta} - \frac{\lambda_3}{t-\Delta} = 0.
\end{equation}
and find that all $\lambda_j$ are real. We then observe that the matrix $(x_1,x_2)$ is invertible and we can evaluate the fixed-point boundary Green function for $\omega=0$. This yields
\begin{equation}
    G(\omega=0)=\frac{2\lambda_1}{t+\Delta}\tau_{z},
\end{equation}
which is nonsingular indicating that there are no zero-energy end states.

\subsubsection{Topological index}

The Kitaev chain with the Hamiltonian given in Eq.~(\ref{eq:H_Kitaev}) belongs to symmetry class BDI, as it has a chiral symmetry operator $\mathcal{U}_S = \tau_x$. For 1D systems, class BDI is characterized by a topological $\mathbb{Z}$-invariant which is given by
\begin{equation}
    \mathcal{Q} = \nu(\lim_{\omega\to 0} \mathcal{U}_S \frac{\mathbb{I}-iVGV^{\dagger}}{\mathbb{I}+iVGV^\dagger}) - \frac{M}{2}.
\end{equation}
Here, $M$ is even and $\nu(F)$ counts the number of negative eigenvalues of a matrix $F$. This expression follows from Ref.\ \cite{Fulga2012} in conjunction with the relation (\ref{eq:rG}) between reflection matrix and boundary Green function (see App.~\ref{app:table} for a summary).

The existence of zero-energy boundary modes implies that for $\omega \to 0$, the boundary Green function has the form
\begin{equation}
    G \sim \frac{K}{\omega} 
    \label{eq:asymptotic}
\end{equation}
with a singular Hermitian matrix $K$ whose rank corresponds to the number of zero modes. This general form for symmetry class BDI is consistent with the expressions (\ref{eq:fpgfkitaevtop}) and (\ref{eq:G34}) which we found above for the Kitaev chain. The chiral symmetry implies that $\mathcal{U}_S$ anticommutes with the Hamiltonian, so that $[U_S,VKV^\dagger]=0$ (see App.~\ref{app:table} for more details). Thus, we can choose a basis in which the Hermitian matrices $\mathcal{U}_S$ and $VKV^{\dagger}$ are simultaneously diagonalized and order the eigenvalues such that
\begin{equation}
    \mathcal{U}_{S}=\diag\left(\mathbb{I}_{M/2},-\mathbb{I}_{M/2}\right)
\end{equation}
with $\mathbb{I}_{n}$ the identity matrix of dimension $n$, and
\begin{equation}
    VKV^{\dagger} = \diag\left(D_{+,p},0_{M/2-p},D_{-,q},0_{M/2-q} \right)
\end{equation}
with $D_{\pm,p}$ a diagonal matrix containing the nonzero eigenvalues of $VKV^{\dagger}$. Here, $p$ ($q$) denotes the number of zero-energy  modes with positive (negative) chirality. We also use the notation $0_{n}$ for an $n\times n$ zero matrix.

Using the asymptotic form of the boundary Green function in Eq.~(\ref{eq:asymptotic}), we thus find
\begin{align}
    F &= \lim_{\omega\to 0} \mathcal{U}_S \frac{\mathbb{I}-iVGV^{\dagger}}{\mathbb{I}+iVGV^\dagger}  \nonumber \\
    &=\diag\left(-\mathbb{I}_{p},\mathbb{I}_{M/2-p},\mathbb{I}_{q},-\mathbb{I}_{M/2-q}\right),
\end{align}
and therefore
\begin{equation}
    \mathcal{Q} = \nu(F) - M/2 = p-q.
\end{equation}
Thus, the topological index of symmetry class BDI can be computed as the difference between the numbers of zero-energy modes with positive and negative chiralities.

In the absence of zero modes, $\lim_{\omega\to 0}G(\omega)=G(0)$ is generically an invertible, full-rank matrix. Then, chiral symmetry implies $\{\mathcal{U}_{S},VG(0)V^\dagger\}=0$. (Notice that unlike for the singular boundary Green function in the topological phase, the leading contribution to $G$ for $\omega\to 0$ is independent of, and thus even in, $\omega$.) Hence, we can choose a basis in which
\begin{equation}
    \mathcal{U}_{S} = \sigma_{x} \otimes \mathbb{I}_{M/2}
\end{equation}
and 
\begin{equation}
    VG(0)V^{\dagger} =  \sigma_{z} \otimes D_{M/2},
\end{equation}
where the diagonal $M/2\times M/2$ matrix $D_{M/2}$ collects eigenvalues of $VG(0)V^{\dagger}$. We obtain the matrix
\begin{align}
    F &= \lim_{\omega\to 0} \mathcal{U}_S \frac{\mathbb{I}-iVGV^{\dagger}}{\mathbb{I}+iVGV^\dagger}  \nonumber \\
    &=(\sigma_{x}\otimes\mathbb{I}_{M/2})\frac{\mathbb{I}_{M}-i(\sigma_{z}\otimes D_{M/2})}{\mathbb{I}_{M}+i(\sigma_{z}\otimes D_{M/2})}.
\end{align}
As $\{\sigma_{z}\otimes \mathbb{I}_{M/2},F\}=0$, the spectrum of $F$ must be symmetric with respect to zero. Hence, we conclude $\mathcal{Q} = 0$ as expected. 

For the Kitaev chain in the topological phase, the asymptotic form of the boundary Green function is $G \sim (1\pm \tau_x)/\omega$. This leads to  the topological invariant $\mathcal{Q} = \pm 1$, corresponding to the existence of one Majorana zero mode with chirality $\pm 1$. When this model is in its trivial phase,  we have $G \sim \tau_z$ and the topological invariant becomes $\mathcal{Q} = 0$.

In realistic 1D topological superconductors, the chiral symmetry will typically be absent. This lowers the symmetry class from BDI to D, which has a $\mathbb{Z}_2$ topological invariant. In this case, there can be at most one zero-energy boundary mode when the system is in the topological phase. Then, the asymptotic Green function $G \sim K/\omega$ involves a matrix $K$ of rank one. The topological invariant for this case is (see App.~\ref{app:table} for details)
\begin{equation}
    \mathcal{Q} = \det\left( \lim_{\omega \to 0}\frac{\mathbb{I} - iVGV^{\dagger}}{\mathbb{I}+iVGV^{\dagger}} \right),
\end{equation}
which takes the value $+1$ in the trivial phase and $-1$ in the topological phase. 

In the presence of one boundary zero mode, diagonalizing the matrix $VKV^{\dagger}$ results in only one non-zero eigenvalue. In this diagonal basis, one finds
\begin{equation}
    \lim_{\omega \to 0}\frac{\mathbb{I} - iVGV^{\dagger}}{\mathbb{I}+iVGV^{\dagger}}=\diag(-1,\mathbb{I}_{M-1}). 
\end{equation}
which leads to $\mathcal{Q}=-1$.

In the trivial phase, $\lim_{\omega\to 0}G(\omega)=G(0)$ is generically a full rank matrix. Because of particle-hole symmetry (see App.~\ref{app:table} for further details), the spectrum of $VG(0)V^{\dagger}$ is symmetric about zero, which leads to $\mathcal{Q}=1$. 

As an aside, this implies that the topological invariant for class D signals the presence or absence of Majoranas in the Kitaev chain despite its being in class BDI. 

\subsection{Chern insulator}

\subsubsection{Model and boundary Green function}

In this section, we consider a Chern insulator as an example of a 2D topological phase. A simple model Hamiltonian for a Chern insulator on a square lattice is given by \cite{Bernevig2013book}
\begin{equation}
    H=\sin k_{x}\sigma_{x}+\sin k_{y}\sigma_{y}+B(2-M-\cos k_{x}-\cos k_{y})\sigma_{z}.
    \label{eq:H_CI}
\end{equation}
The corresponding spectrum is gapped except for $M=0,2,4$, and the model exhibits topological phases for $0<M<2$ --  with edge states around $k=0$ -- and for $2<M<4$ -- with edge states around $k = \pi$. 

To compute the boundary Green function, we rewrite the Hamiltonian in real space along the $x$-direction (and retain the momentum representation in the $y$-direction with periodic boundary conditions). This yields
\begin{align}
    H(k_{y})&=\sum_{n=0}^{N}\bigg[\frac{1}{2i}c_{n+1}^{\dagger}\sigma_{x}c_{n}-\frac{B}{2}c_{n+1}^{\dagger}\sigma_{z}c_{n} + {\rm h.c.}  
     \nonumber \\
       & +c_{n}^{\dagger}(\sin k_{y}\sigma_{y}+B\Lambda(k_{y})\sigma_{z})c_{n} \bigg]
\end{align}
with $\Lambda(k_{y})=2-M-\cos k_{y}$. We can then readily read off 
\begin{gather}
    g^{-1}(k_y,\omega)=\omega-\sin k_{y}\sigma_{y}-B\Lambda(k_{y})\sigma_{z}\\
    V=\frac{1}{2i}\sigma_{x}-\frac{B}{2}\sigma_{z}.
\end{gather}

It is useful to compute the fixed-point boundary Green function in a perturbation theory in $(\omega-\sin k_y\sigma_y)$, starting with
\begin{equation}
    T=T_{0}+A(\omega-\sin k_{y}\sigma_{y}),\quad A=\left(\begin{array}{cc}
      (V^\dagger)^{-1} & 0\\
        0 & 0
    \end{array}\right),
\end{equation}
where $T_{0}=T(\sin k_{y}=0,\omega=0)$. For $\omega-\sin k_{y}\sigma_{y}=0$, the eigenvalues satisfy
\begin{equation}
    \lambda^{2}(1\pm B)\mp2B\Lambda\lambda-(1\mp B)=0.
\end{equation}
We denote the roots for the upper sign as $\lambda_{1,2}$ and for the lower sign as $\lambda_{3,4}$, and fix the labeling by imposing $\lambda_{1}\lambda_{4}^{*}=1=\lambda_{2}\lambda_{3}^{*}$. The corresponding eigenvectors are 
\begin{equation}
    z_{1}=z_{2}=\left(\begin{array}{c}
        1\\
        i
    \end{array}\right),\quad z_{3}=z_{4}=\left(\begin{array}{c}
        1\\
        -i
    \end{array}\right).
\end{equation}
Similar to the Kitaev chain, we expect a topological phase when $|\lambda_1|$ and $|\lambda_2|$ are either both larger or both smaller than unity. The relevant parameter ranges can again be deduced from the quantity 
\begin{equation}
 (\lambda_1^2 -1)(\lambda_2^2 -1) =\frac{4B^2}{(1+B)^2}(1-\Lambda^2).
\end{equation}
Thus, the topological phase requires $\Lambda^2<1$, which implies $\abs{\lambda_{1,2}}>1$ for $B<0$ and $\abs{\lambda_{3,4}}>1$ for $B>0$. Using that $\cos k_y=\pm 1$ when $\sin k_y=0$, we recover that the model is topological for $0<M<4$. 

We can now evaluate the fixed-point boundary Green function in perturbation theory by following essentially the same steps as for the Kitaev chain. When $\abs{\lambda_{1,2}}>1$, a straight-forward calculation yields
\begin{equation}
    G(k_y,\omega) = \frac{(1+\sigma_{y})}{2(B+1)(\omega+\sin k_{y})a_{12}}
\end{equation}
with
\begin{align}
    a_{12} &=
    \frac{1}{\left[-(B+1)\lambda_{1}^{*}+(B-1)\lambda_{4}\right](\abs{\lambda_{1}}^{2}-1)(\lambda_{2}-\lambda_{4})} \nonumber \\
    &+\frac{1}{\left[-(B+1)\lambda_{2}^{*}+(B-1)\lambda_{3}\right](\abs{\lambda_{2}}^{2}-1)(\lambda_{1}-\lambda_{3})}.
\end{align}
Similarly, when $\abs{\lambda_{3,4}}>1$, we have
\begin{equation}
    G(k_y,\omega) = \frac{(1-\sigma_{y})}{2(B-1)(\omega-\sin k_{y})a_{34}}
\end{equation}
where
\begin{align}
  a_{34} &= \frac{1}{(\abs{\lambda_3}^2 -1)(\lambda_4-\lambda_1)\left[(1-B)\lambda_4^* + (B+1)\lambda_1\right]}
  \nonumber \\
  &+\frac{1}{(\abs{\lambda_4}^2 -1)(\lambda_3-\lambda_2)\left[(1-B)\lambda_3^* + (B+1)\lambda_2\right]}.
\end{align}
Thus, we find that when $0<M<2$, the system has a gapless chiral boundary mode near $k_{y}= 0$
with dispersion $\omega = \Sgn(B)\sin k_{y}$. When $2<M<4$, the gapless chiral boundary mode occurs around $k_{y}= \pi$ with dispersion $\omega = \Sgn(B)\sin k_{y}$. 

In the nontopological phase for $M<0$ or $M>4$, we have $\Lambda^2 >1$ and can thus choose $\abs{\lambda_1}, \abs{\lambda_3}>1$. As for the Kitaev chain, the $\lambda_j$ are all real and obey the relation
\begin{equation}
    \frac{\lambda_1}{1-B} + \frac{\lambda_3}{1+B} = 0.
\end{equation}
The boundary Green function is nonsingular and can hence be evaluated approximately for $\omega-\sin k_y=0$. This yields
\begin{equation}
    G(k_{y}\to 0,\pi,\omega\to 0) = - \frac{2\lambda_1}{1-B}\sigma_z.
\end{equation}

\subsubsection{Topological index}

The Chern insulator with Hamiltonian (\ref{eq:H_CI}) belongs to class A. The topological invariant for a two-dimensional system in this symmetry class takes the form \cite{Fulga2012} (see also App.~\ref{app:table})
\begin{equation}
    \mathcal{Q} = \frac{1}{2\pi i} \int _{0}^{2\pi}dk\,\frac{d}{dk} \Tr \log  r(k),
\end{equation}
where the zero-energy reflection matrix is expressed in terms of the boundary Green function as
\begin{equation}
    r(k_{y}) = \lim_{\omega\to 0} \frac{\mathbb{I}-iVG(k_{y},\omega)V^{\dagger}}{\mathbb{I}+iVG(k_{y},\omega)V^\dagger}.
\end{equation}
In line with Laughlin's argument \cite{Laughlin1981} and the familiar expression for quantum pumping of charge \cite{Brouwer1998}, this can be viewed as the pumped charge in units of the electron charge when threading the effectively cylinder-shaped sample (periodic boundary conditions in the $y$-direction) by one flux quantum. 

When $0<M<2$, the boundary Green function diverges at $k_{y}=0$ and has rank one, 
\begin{equation}
    G(k_y,\omega \to 0) \sim \frac{1+\sigma_y}{k_y}.
\end{equation}
Away from $k_y=0$ the boundary Green function is invertible and finite. This yields a winding number $\mathcal{Q} = 1$.

When $2<M<4$, the boundary Green function diverges  at $k_{y} = \pi$ and has rank one,
\begin{equation}
    G(k_y,\omega \to 0) \sim -\frac{1+\sigma_y}{k_y - \pi}.
\end{equation}
Away from $k_y = \pi$ the boundary Green function is well behaved. We thus obtain the opposite winding number $\mathcal{Q}=-1$.

When $M<0$ or $M>4$, the boundary Green function is well behaved at all $k_y$ and the winding number is zero.

\section{Higher-order topological insulators  \label{sec:inhomogeneous}}

\subsection{Model}

Recent work introduced higher-order topological insulators \cite{Benalcazar2016}. In 2D, these phases have gapped edges but gapless corner states. In this section, we show that our approach readily accommodates such systems and provides a rather transparent picture. 

\begin{figure}[t!]
    \includegraphics[width=0.45\textwidth]{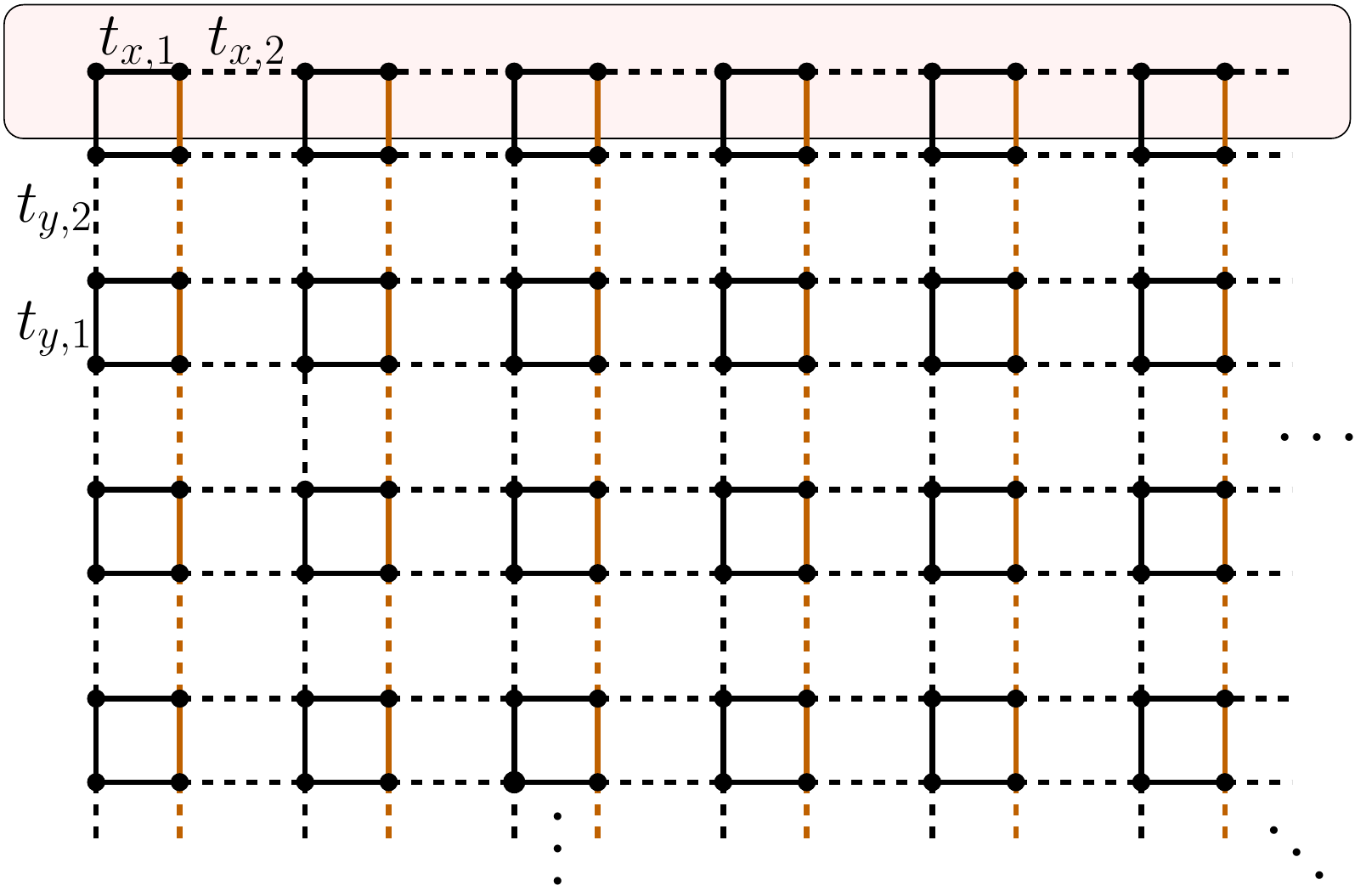}
    \caption{Tight-binding model for a second order topological insulator on a square lattice. Solid and dashed lines in the $x$ and $y$ directions denote alternating nearest neighbor hoppings $t_{x,1}$, $t_{x,2}$ and $t_{y,1}$, $t_{y,2}$. The hopping amplitudes for the bonds colored in brown include a relative minus sign. The effective Hamiltonian of the highlighted boundary parallel to the $x$-direction is topologically equivalent to an SSH model.}
    \label{fig:2D}
\end{figure}

The model investigated in Ref.~\cite{Benalcazar2016} takes the form
\begin{align}
  H &= \sum_{n,j} \left( t_{x,1}c_{2n,j}^{\dagger}c_{2n-1,j} + t_{x,2} c_{2n+1,j}^{\dagger}c_{2n,j}  \right) \nonumber \\
  &- \sum_{i,m}  (-1)^{i}\left(t_{y,1}c_{i,2m}^{\dagger}c_{i,2m-1} + t_{y,2} c_{i,2m+1}^{\dagger}c_{i,2m}  \right).
  \label{eq:BBH}
\end{align}
This square-lattice Hamiltonian has dimerized hopping amplitudes in the $x$ and $y$-directions and an added $\pi$-flux per plaquette, as illustrated in Fig.~\ref{fig:2D}. The hopping amplitudes $t_{\alpha,j}$ ($\alpha=x,y$ and $j=1,2$) are assumed to be real. As a consequence of the $\pi$ flux, the model has an insulating bulk which makes it amenable to our boundary Green function approach.  

While the edges of this model are gapped, there are zero-energy corner states \cite{Benalcazar2016}. In the presence of $C_4$ symmetry, these corner states can be characterized by a quadrupole moment which
is a $\mathbb{Z}_2$ topological index. In the absence of  $C_4$ symmetry, there is a $\mathbb{Z}_2\times \mathbb{Z}_2$ topological index which can be associated with the edge polarizations. 

The main observation underlying the current section is that even if the fixed-point boundary Green function is gapped, it defines a low-energy boundary Hamiltonian in a natural manner. This boundary Hamiltonian can be either trivial or topological. Gapless corner states (or corresponding generalizations to higher dimensions) appear if this boundary Hamiltonian is topological. 

\subsection{Effective boundary Hamiltonian \label{sec:2nd_order}}

We now derive the fixed-point boundary Green function of this model. The dimerization of the hopping amplitudes makes this a model with period $a=2$, with $T=M_2M_1$ being the product of two matrices. As described in Sec.\ \ref{sec:GF_TM}, we can deduce the fixed-point boundary Green function from the eigenvectors and eigenvalues of $T$. 

Consider a system with $2N\times 2N$ sites and focus on the Green function for a boundary parallel to the $x$-direction (in which we thus assume periodic boundary conditions). Introducing a two-component spinor $\psi_{n,j}^T=(c_{2n-1,j}^T,c_{2n,j}^T)$ and its Fourier transform
\begin{equation}
  \psi_{n,j} = \frac{1}{\sqrt{N}}\sum_{k} e^{ink}\psi_j(k),
\end{equation}
the Hamiltonian can be rewritten as
\begin{align}
  H=&\sum_{k} \left\{ \sum_{j=1}^{2N}\psi_{j}^{\dagger}(k)h_{j}(k)\psi_{j}(k)  \right.\nonumber \\
  &+ \left. \sum_{j=1}^{2N-1}\left[\psi_{j+1}^{\dagger}(k)V_{j}\psi_{j}(k)+{\rm h.c.}\right]\right\}.
\end{align}
Here, we define 
\begin{equation}
  h_{n}(k)=t_{x,1}\tau_x+t_{x,2}\left(e^{-ik}\tau_{+} + e^{ik}\tau_-\right),
\end{equation}
and
\begin{equation}
  V_{j}=\begin{cases}
    t_{y,1}\tau_{z} & \quad{\rm odd}\ j\\
    t_{y,2}\tau_{z} & \quad{\rm even}\ j
    \end{cases}.
\end{equation}
The matrices $\tau_{x,y,z}$ are Pauli matrices in the sublattice space of even and odd sites along the $x$-direction, with $\tau_{\pm} =(\tau_x \pm i\tau_y)/2$.

With these ingredients, we can construct the transfer matrix $T= M_2 M_1$ using Eq.\ (\ref{eq:Mj}). As the edge is gapped, we can set $\omega =0$ from the outset. This yields
\begin{widetext}
\begin{equation}
  T = \left(\begin{array}{cccc}
      -\frac{t_{x,1}^2+t_{x,2}^2+t_{y,1}^2+2t_{x,1}t_{x,2}\cos k}{t_{y,1}t_{y,2}} & 0 & 0 & \frac{t_{x,1}}{t_{y,1}}+\frac{t_{x,2}}{t_{y,1}}e^{-ik}\\
0 & -\frac{t_{x,1}^2+t_{x,2}^2+t_{y,1}^2+2t_{x,1}t_{x,2}\cos k}{t_{y,1}t_{y,2}} & -\frac{t_{x,1}}{t_{y,1}}-\frac{t_{x,2}}{t_{y,1}}e^{ik} & 0\\
0 & -\frac{t_{x,1}}{t_{y,1}}-\frac{t_{x,2}}{t_{y,1}}e^{-ik} & -\frac{t_{y,2}}{t_{y,1}} & 0\\
\frac{t_{x,1}}{t_{y,1}}+\frac{t_{x,2}}{t_{y,1}}e^{ik} & 0 & 0 &-\frac{t_{y,2}}{t_{y,1}} 
\end{array}\right).
\end{equation}
\end{widetext}
The matrix $T$ decomposes into two independent $2\times 2$ blocks, which we can treat separately. Both blocks have the same eigenvalues $\lambda$ (with $|\lambda|>1$) and $1/\lambda^*$. We find that $\lambda$ fulfills the equation
\begin{equation}
  \lambda^{2}+\lambda\frac{t_{x,1}^{2}+t_{x,2}^{2}+t_{y,1}^{2}+t_{y,2}^{2}+2t_{x,1}t_{x,2}\cos k}{t_{y,1}t_{y,2}}+1=0.
\end{equation}
Evaluating the corresponding eigenvectors, we can then compute the fixed-point boundary Green function based on Eq.\ (\ref{eq:Ninf}). This yields 
\begin{align}
  G(0,k) =& \frac{1}{t_{y,2}(t_{y,2}+\lambda t_{y,1})} \nonumber \\
  &\times
  \left(\begin{array}{cc} 0 & t_{x,1}+t_{x,2}e^{-ik}\\ t_{x,1}+t_{x,2}e^{ik} & 0
\end{array}\right),
\end{align}
which is invertible and purely real, reflecting the fact that this describes the gapped edge of a gapped bulk. 

We can now define an effective boundary Hamiltonian 
\begin{align}
  H_{\rm eff} &= -\left[G(0,k)\right]^{-1} \nonumber
  \\
  &=f_{k}\left(\begin{array}{cc}
0 & t_{x,1}+t_{x,2}e^{-ik}\\
t_{x,1}+t_{x,2}e^{ik} & 0
\end{array}\right)
\end{align}
from the fixed-point boundary Green function. Here, the prefactor 
\begin{equation}
  f(k) = \frac{t_{y,2}(t_{y,2}+\lambda t_{y,1})}{t_{x,1}^2 + t_{x,2}^2 + 2t_{x,1}t_{x,2}\cos k}
\end{equation}
has a fixed sign as a function of $k$. It is our central observation that up to the prefactor $f_k$, this is just the Hamiltonian of the SSH model (\ref{eq:SSH}). Thus, the boundary Hamiltonian is topologically equivalent to the bulk Hamiltonian of the SSH model, with $t_1$ and $t_{2}$ replaced by $t_{x,1}$ and $t_{x,2}$. 

A similar analysis can be performed for the edge parallel to the $y$-direction.  In view of the fact that the $\pi$-flux can be equally included via the hopping amplitudes in the $x$-direction, the calculation is entirely equivalent and just results in exchanging the subscripts $x$ and $y$. Thus, the edge along the $y$-direction can also be characterized by a $\mathbb{Z}_2$ invariant. Altogether, model is thus characterized by a $\mathbb{Z}_2 \times \mathbb{Z}_2$ invariant, which encodes the polarizations of the edges in the two directions \cite{Benalcazar2016}.

\begin{figure}[t!]
    \includegraphics[width=0.45\textwidth]{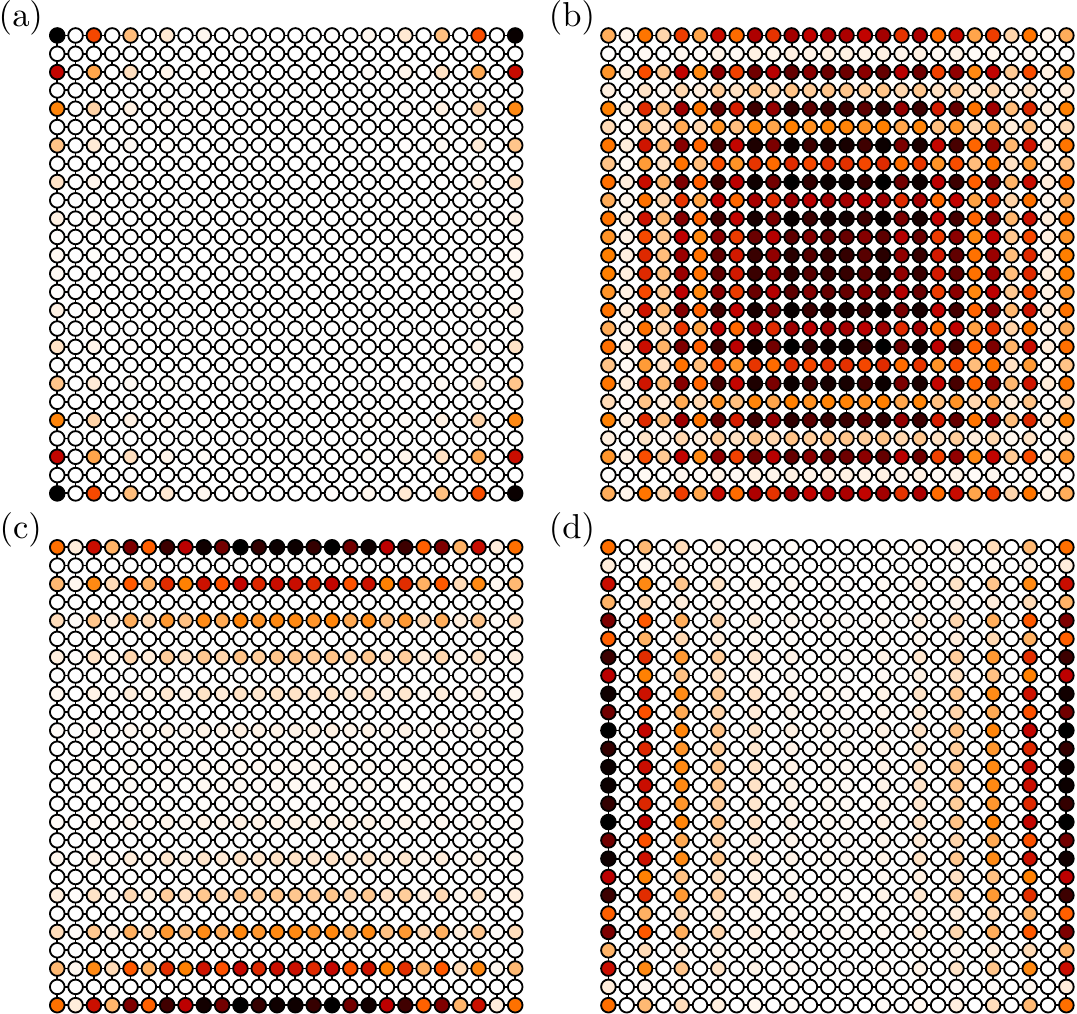}
    \caption{Wave function (modulus squared) of the single-particle excitation closest to zero energy for the second-order topological insulator in Eq.~(\ref{eq:BBH}), evaluated on a finite-size lattice. Darker color indicates larger magnitude of the wave functions. The parameters are chosen as: (a) $t_{x,1}=0.7$, $t_{y,1}=0.6$;(b) $t_{x,1}=1.2$, $t_{y,1}=1.5$; (c)  $t_{x,1}=1.2$, $t_{y,1}=0.7$; and (d)  $t_{x,1}=0.7$, $t_{y,1}=1.5$. We choose $t_{x,2}=t_{y,2}=1$ in all cases. }
    \label{fig:wavefunctions}
\end{figure}

The various cases can be illustrated by plotting the wave function (modulus squared) of the excitation which is closest to zero energy, see Fig.\ \ref{fig:wavefunctions}. When the boundary Hamiltonians for both the $x$- and the $y$-edges are in the topological phase, the model exhibits corner states [panel (a)]. The excitation corresponds to a pure bulk state when both edge Hamiltonians are in the nontopological phase [panel (b)]. When only one of the two boundary Hamiltonians is in the topological phase, the excitations are localized at the boundary in the topological direction, but delocalized in the other one [panels (c) and (d)].

\section{Conclusion \label{sec:conclusion}}

The boundary Green function is a natural quantity to characterize topological insulators and superconductors as it directly encodes the boundary modes and hence the topological phase diagram as well as the topological indices. In this paper, we have developed a systematic approach to compute boundary Green functions of topological insulators and superconductors by relying on a recursive approach. We show for several familiar models that our approach can be applied in a rather straight-forward manner. 

Though simpler, our recursive approach has some similarities with a real-space renormalization group approach. For the noninteracting fermion lattice problems discussed in this paper, we find that the recursion relation can be solved explicitly in terms of the transfer matrix which greatly facilitates analytical calculations.   
When attaching leads to the sample, we can relate the boundary Green function to the reflection matrix. In view of earlier work on deriving the periodic table of topological phases of free fermions from the reflection matrix \cite{Fulga2012}, this provides us with explicit expressions for the topological indices in terms of the boundary Green functions. 

As an interesting application, we show that for insulators with trivial bulk topology, the boundary Green function defines a boundary Hamiltonian in a natural manner. Remarkably, even if the bulk is nontopological, this boundary Hamiltonian can still be topologically nontrivial which, say in 2D, is reflected in gapless corner states. We show this explicitly for a model which was recently proposed. This not only provides an intuitive approach to these recently introduced topological systems, but also has some potential to serve as a starting point for a more systematic investigation of these higher-order topological insulators. 

While we restricted ourselves to clean and noninteracting models, the recursive approach should be extendable to include disorder and possibly interactions. For instance, in the presence of disorder, the recursion for the boundary Green function remains valid. While for a fixed disorder configuration, the boundary Green function would no longer approach a fixed-point Green function in the thermodynamic limit, the distribution of boundary Green functions over the disorder ensemble should still have such a fixed point. 

\acknowledgments 
We thank Piet Brouwer and Andrei Bernevig for a critical reading of the manuscript, and acknowledge financial support by the Deutsche Forschungsgemeinschaft CRC 183. Yimu Bao thanks the Dahlem Center for hospitality during a research internship. 

\appendix

\section{Boundary Green function and reflection matrix\label{app:BGF_R}}

In this appendix, we present an alternative derivation of the relation (\ref{eq:rG}) between reflection matrix and boundary Green function. 

We start by noting that if we multiply $\psi_n^\dagger$ from the left in Eq.~(\ref{eq:SEQ}), we find that $\mathrm{Im}\psi_{n}^{\dagger}V_{n}^{\dagger}\psi_{n+1}$ is independ of $n$. Thus, we can define the current operator
\begin{equation}
  \hat{\mathcal{I}}_n =  i\Omega_n= i\left(\begin{array}{cc}
    0 & -V_{n}\\
    V_{n}^{\dagger} & 0
  \end{array}\right),
\end{equation}
so that the current  $\mathcal{I} = \Psi(n)^\dagger \hat{\mathcal{I}}_n \Psi(n) $ becomes independent of $n$. Here, we use the notation $\Psi(n) = [\psi(n+1)^T, \psi(n)^T]^T$.

It is more convenient to have a current operator which is independent of cell index $n$. To this end, we introduce the transformation
\begin{equation}
    \Psi(n)\to O_{n}\Psi(n),\quad \hat{\mathcal{I}}_n \to \left(O_{n}^{\dagger}\right)^{-1}\hat{\mathcal{I}}_n O_{n}^{-1}=\sigma_{y}\otimes \mathbb{I}
\end{equation}
with $O_n =  \diag(1,V_{n})$. Here, the Pauli matrices $\sigma_{x,y,z}$ act in the two-component space of $\Psi(n)$. In the transformed basis, the matrix $M_n$ defined in Eq.~(\ref{eq:Mj}) becomes
\begin{equation}
  M_n \to O_{n} M_n O_{n-1}^{-1}
  =\left(\begin{array}{cc}
    (V_{n}^{\dagger})^{-1}g_{n}^{-1} & -(V_{n}^{\dagger})^{-1}\\
    V_{n} & 0
  \end{array}\right),
\end{equation}
so that the relation $\Psi(n+1) = M_n \Psi(n)$ remains unchanged. 

We can further diagonalize the current operator by introducing a unitary transformation 
\begin{equation}
    P=\frac{1}{\sqrt{2}}\left(\begin{array}{cc}
        1 & -i\\
        1 & i
    \end{array}\right),
\end{equation}
such that
\begin{equation}
    \Psi(n)\to P\Psi(n),\quad M_{n}\to PM_{n}P^{\dagger}.
\end{equation}
Then, the current operator becomes
\begin{equation}
    \hat{\mathcal{I}} = \sigma_z \otimes \mathbb{I}.
\end{equation}
In this new basis, the upper and lower components (of length $M$ each) of $\Psi(n)$ describe the right- and left-moving states at unit cell $n$.

If we attach normal metal leads from left and right to the first and the last unit cell of the quasi-1D system of $N$ cells, we can define the scattering matrix 
\begin{equation}
    S_{N}=\left(\begin{array}{cc}
        r'_{N} & t_{N}\\
        t'_{N} & r_{N}
    \end{array}\right),
\end{equation}
where the $2\times2$ block structure reflects the two leads,  $r_{N}, r'_{N}$  are reflection matrices and $t_{N},t'_{N}$  are transmission matrices. The scattering matrix relates the outgoing states on left and right to the incoming states. After the basis transformations, we have
\begin{equation}
    \Psi(n)=\left(\begin{array}{c}
        \psi(n+1)-iV_{n}\psi(n)\\
        \psi(n+1)+iV_{n}\psi(n)
    \end{array}\right).
\end{equation}
Here the upper and lower components are outgoing (incoming) and incoming (outgoing) states on the right (left).  

In the new basis, the transfer matrix can be written as \cite{Nazarov}
\begin{equation}
    \mathcal{M}_{N}=\left(\begin{array}{cc}
        t'_{N}-r_{N}t_{N}^{-1}r'_{N} & r_{N}t_{N}^{-1}\\
        -t_{N}^{-1}r'_{N} & t_{N}^{-1}
    \end{array}\right).
\end{equation}
If we add one unit cell to the system, we have $\mathcal{M}_{N+1}=M_{N+1}\mathcal{M}_{N}$,
where $M_{N}$ in the new basis becomes
\begin{widetext}
\begin{equation}
    M_N = \frac{1}{2}\left(\begin{array}{cc}
        (V_{N}^{\dagger})^{-1}g_{N}^{-1}-iV_{N}-i(V_{N}^{\dagger})^{-1} & (V_{N}^{\dagger})^{-1}g_{N}^{-1}-iV_{N}+i(V_{N}^{\dagger})^{-1}\\
        (V_{N}^{\dagger})^{-1}g_{N}^{-1}+iV_{N}-i(V_{N}^{\dagger})^{-1} & (V_{N}^{\dagger})^{-1}g_{N}^{-1}+iV_{N}+i(V_{N}^{\dagger})^{-1}
    \end{array}\right)
\end{equation}
by straightforward algebra. We then find the recurrence relations 
\begin{gather}
    r_{N+1}t_{N+1}^{-1} =
    \frac{1}{2}\left[(V_{N+1}^{\dagger})^{-1}g_{N+1}^{-1}-iV_{N+1}\right](r_{N}+\mathbb{I})t_{N}^{-1}-\frac{i}{2}(V_{N+1}^{\dagger})^{-1}(r_{N}-\mathbb{I})t_{N}^{-1} \\
    t_{N+1}^{-1}  = \frac{1}{2}\left[(V_{N+1}^{\dagger})^{-1}g_{N+1}^{-1}+iV_{N+1}\right](r_{N}+\mathbb{I})t_{N}^{-1}-\frac{i}{2}(V_{N+1}^{\dagger})^{-1}(r_{N}-\mathbb{I})t_{N}^{-1}.
\end{gather}
for the reflection and transmission matrices. By adding and subtracting these two equations, we find
\begin{align}
    (r_{N+1}+\mathbb{I})t_{N+1}^{-1} & =(V_{N+1}^{\dagger})^{-1}g_{N+1}^{-1}(r_{N}+\mathbb{I})t_{N}^{-1}-i(V_{N+1}^{\dagger})^{-1}(r_{N}-\mathbb{I})t_{N}^{-1}\\
    (r_{N+1}-\mathbb{I})t_{N+1}^{-1} & =-iV_{N+1}(r_{N}+\mathbb{I})t_{N}^{-1}.
\end{align}
From the last equation, we have
\begin{equation}
    t_{N}^{-1}=i(r_{N}+\mathbb{I})^{-1}V_{N+1}^{-1}(r_{N+1}-\mathbb{I})t_{N+1}^{-1}.
\end{equation}
Inserting this expression back into the previous one, we find
\begin{equation}
    (r_{N+1}+\mathbb{I})t_{N+1}^{-1}=i(V_{N+1}^{\dagger})^{-1}\left[g_{N+1}^{-1}(r_{N}+\mathbb{I})-i(r_{N}-\mathbb{I})\right](r_{N}+\mathbb{I})^{-1}V_{N+1}^{-1}(r_{N+1}-\mathbb{I})t_{N+1}^{-1}.
\end{equation}
Multiplying $t_{N+1}$ from the right, we obtain
\begin{equation}
    (r_{N+1}+\mathbb{I})(r_{N+1}-\mathbb{I})^{-1}=i(V_{N+1}^{\dagger})^{-1}\left[g_{N+1}^{-1}-i(r_{N}-\mathbb{I})(r_{N}+\mathbb{I})^{-1}\right]V_{N+1}^{-1}.
\end{equation}
\end{widetext}
Defining
\begin{equation}
    \mathcal{R}_{N}=(r_{N}-\mathbb{I})(r_{N}+\mathbb{I})^{-1},
\end{equation}
we have
\begin{equation}
i\mathcal{R}_{N+1}=V_{N+1}\left[g_{N+1}^{-1}-i\mathcal{R}_{N}\right]^{-1}V_{N+1}^{\dagger}.\label{eq:recurrence}
\end{equation}
Comparing with Eq.~(\ref{eq:Dyson}), we find that the boundary Green function can be written as
\begin{equation}
G_{N}=V_{N}^{-1}i\mathcal{R}_{N}(V_{N}^{\dagger})^{-1}.
\end{equation}
Solving for $r_N$, we recover the relation (\ref{eq:rG}) between the reflection matrix $r_{N}$ and the boundary Green function $G_{N}$.

\section{Topological invariants \label{app:table}}

Single particle Hamiltonians can be classified based on nonspatial and nonunitary symmetries,
i.e., the time-reversal, particle-hole, and chiral symmetries. This defines the ten Altland-Zirnbauer symmetry classes \cite{Altland97}. Ref.~\cite{Fulga2012} discussed the topological classification of the ten symmetry classes in terms of the reflection matrix at the Fermi energy. In this section, we adapt this discussion to the language of the boundary Green functions and provide explicit expressions for the topological indices.

\subsection{Bott periodicity\label{sec:general}}

\subsubsection{Symmetries}

Time-reversal symmetry requires the existence of an antiunitary operator $\hat{\mathcal{T}}=\mathcal{U}_{T}\hat{\mathcal{K}}$ with a unitary matrix $\mathcal{U}_{T}$ and complex conjugation operator  $\hat{\mathcal{K}}$, such that
\begin{equation}
    \hat{\mathcal{T}} O(\v{k}_{\perp}) = O(-\v{k}_{\perp}) \hat{\mathcal{T}}, \quad O=h_n,V_n\quad \forall n.
    \label{eq:TR1}
\end{equation}
Correspondingly, the boundary Green function obeys the relation
\begin{equation}
    \hat{\mathcal{T}} G_{n}(\mathbf{k}_{\perp},\omega) = G_{n}(-\mathbf{k}_{\perp},\omega^*) \hat{\mathcal{T}},
    \label{eq:TR2}
\end{equation}
which can be proven iteratively using the Dyson equation [see Eq.~(\ref{eq:Dyson})]. If we apply the time-reversal operator twice, we obtain 
\begin{equation}
    [\mathcal{U}_{T}^* \mathcal{U}_{T}, h_n]=[\mathcal{U}_{T}^* \mathcal{U}_{T}, V_n] = [\mathcal{U}_{T}^* \mathcal{U}_{T}, V_n^{\dagger}]=0, \quad \forall n.
\end{equation}
By Schur's lemma, we conclude that $\mathcal{U}_{T}^* \mathcal{U}_{T} = \exp(i\alpha) \mathbb{I}$ is a multiple of the identity matrix. Moreover, because of the unitarity of $\mathcal{U}_{T}$, we find $\exp(2 i\alpha)=\pm 1$. Thus, there are two types of time-reversal symmetries, 
\begin{equation}
    \mathcal{U}_{T}^* \mathcal{U}_{T}= \pm \mathbb{I}
    \label{eq:Tsquare}
\end{equation}
or $\mathcal{T}^2=\pm 1$.

Particle-hole symmetry  requires the existence of an antiunitary operator $\hat{\mathcal{C}}=\mathcal{U}_{C}\hat{\mathcal{K}}$ such that
\begin{equation}
    \hat{\mathcal{C}} O(\v{k}_{\perp}) = - O(-\v{k}_{\perp}) \hat{\mathcal{C}}, \quad O=h_n,V_n\quad \forall n.
\end{equation}
Here, $\mathcal{U}_{C}$ is a unitary matrix. The constraint of particle-hole symmetry on the boundary Green function is 
\begin{equation}
    \hat{\mathcal{C}} G_{n}(\mathbf{k}_{\perp},\omega) = -G_{n}(-\mathbf{k}_{\perp},-\omega^*) \hat{\mathcal{C}}.
\end{equation}
Similar to the case of time-reversal symmetry, we find that there are two types of particle-hole symmetries
given by
\begin{equation}
    \mathcal{U}_{C}^* \mathcal{U}_{C}= \pm \mathbb{I}.
    \label{eq:Csquare}
\end{equation}
or $\hat{\mathcal{C}}^2=\pm 1$.

Chiral symmetry requires the existence of a unitary operator $\hat{\mathcal{S}} = \mathcal{U}_{S}$ 
represented by a unitary matrix $\mathcal{U}_{S}$, such that
\begin{equation}
    \{\hat{\mathcal{S}}, h_n\}=\{\hat{\mathcal{S}}, V_n\} = \{\hat{\mathcal{S}}, V_n^{\dagger}\}=0, \quad \forall n.
\end{equation}
Because of chiral symmetry, the boundary Green function fulfills
\begin{equation}
    \mathcal{U}_{S} G_{n}(\mathbf{k}_{\perp},\omega) = -G_{n}(\mathbf{k}_{\perp},-\omega) \mathcal{U}_{S}.
    \label{eq:GGGSS}
\end{equation}
Applying chiral symmetry twice, we find that $\mathcal{U}_{S}^2=\exp(i\alpha)\mathbb{I}$. Redefining $\mathcal{U}_{S} \to \mathcal{U}_{S} \exp(-i\alpha/2)$, we have 
\begin{equation}
    \mathcal{U}_{S}^2 = \mathbb{I}, \label{eq:Ssquare}
\end{equation}
which implies $\mathcal{U}_{S}=\mathcal{U}_{S}^{\dagger}$. When both time-reversal and particle-hole symmetry exist, the system also has chiral symmetry with $\mathcal{U}_S =\mathcal{U}_T \mathcal{U}_C^*$. If either time-reversal or particle-hole symmetry is absent, chiral symmetry is also absent. However, chiral symmetry can exist when both time-reversal and particle-hole symmetry are absent.

The absence or presence of time-reversal ($\hat{\mathcal{T}}$), particle-hole ($\hat{\mathcal{C}}$), and chiral symmetries ($\hat{\mathcal{S}}$) of the various kinds according to Eqs.~(\ref{eq:Tsquare}), (\ref{eq:Csquare}), and (\ref{eq:Ssquare}) leads to the ten Altland-Zirnbauer symmetry classes, as summarized in the first two columns of Table \ref{tab:tenfold}.

\subsubsection{Dimensional reduction}

In all symmetry classes with a chiral symmetry, the reflection matrix (the subscript $N$ is omitted for simplicity) satisfies the constraint
\begin{equation}
    \mathcal{U}_{S} r(\mathbf{k}_{\perp},\omega) = r^{\dagger}(\mathbf{k}_{\perp},-\omega) \mathcal{U}_{S},
\end{equation}
which follows from Eqs.\ (\ref{eq:rG}) and (\ref{eq:GGGSS}). This implies that in the limit $\omega\to 0$, the matrix $\mathcal{U}_{S}r$ is hermitian where we use the shorthand  $r = \lim_{\omega\to 0}r(\omega)$. This enables one to define an effective Hamiltonian in $d-1$ dimensions as \cite{Fulga2012}
\begin{equation}
    H_{d-1} 
    =  \mathcal{U}_{S}r  =
    \lim_{\omega\to 0}
      \mathcal{U}_{S}
    \frac{\mathbb{I} - iVGV^{\dagger}}{\mathbb{I} + iVGV^{\dagger}}
    \label{eq:effH_chiral}.
\end{equation}
This Hamiltonian is again gapped as it has eigenvalues $\pm 1$ and has no chiral symmetry.

In the absence of chiral symmetry, the eigenvalues of the reflection matrix $r$ at the Fermi energy are in general complex. One can then construct an effective Hamiltonian in $d-1$ dimensions by doubling the degrees of freedom, \cite{Fulga2012}
\begin{align}
    H_{d-1}&=\left(\begin{array}{cc}
        0 & r\\
        r^{\dagger} & 0
    \end{array}\right) \nonumber \\
    &=\lim_{\omega\to0}\left(\begin{array}{cc}
        0 & \frac{\mathbb{I}-iVGV^{\dagger}}{\mathbb{I}+iVGV^{\dagger}}\\
    \frac{\mathbb{I}+iVGV^{\dagger}}{\mathbb{I}-iVGV^{\dagger}} & 0
    \end{array}\right).
    \label{eq:effH_nonchiral}
\end{align}
Because of the identity
\begin{equation}
    \{H_{d-1}, \sigma_z\otimes \mathbb{I} \}=0,
\end{equation}
the effective Hamiltonian acquires a chiral symmetry with $\mathcal{U}_{S} = \sigma_z\otimes \mathbb{I}$. 

{\em Classes A and AIII.---}Hamiltonians in classes A and AIII are thus transformed into one another under this process of dimensional reduction. This is referred to as complex Bott periodicity (mod $2$). 

Now we take time-reversal and particle-hole symmetries into account, in order to reproduce the real Bott periodicity for the remaining eight classes of the ten-fold table. 

{\em Classes AI and AII.---}We first focus on classes AI and AII with only time-reversal symmetry. Let us denote $r(\mathbf{k}_{\perp})=\lim_{\omega\to 0}r(\mathbf{k}_{\perp},\omega)$. By Eqs.~(\ref{eq:TR1}) and (\ref{eq:TR2}), time-reversal symmetry restricts the reflection matrix at the Fermi energy to obey
\begin{equation}
    r^{\dagger}(\mathbf{k}_{\perp})\mathcal{U}_{T} = \mathcal{U}_{T} r^*(-\mathbf{k}_{\perp}).  
    \label{t1}
\end{equation}
Taking the hermitian conjugate and multiplying by $\mathcal{U}_T$ from both sides, we obtain
\begin{equation}
     r(\mathbf{k}_{\perp})\mathcal{U}_{T}=  \mathcal{U}_{T} r^{T}(\mathbf{-k}_{\perp}) .  
\label{t2}
\end{equation}

According to Eq.~(\ref{eq:effH_nonchiral}), the effective Hamiltonian $H_{d-1}$ has chiral symmetry. In addition, we can confirm using Eqs.\ (\ref{t1}) and (\ref{t2}) that is also acquires time-reversal symmetry with $\hat{\mathcal{T}}= (\sigma_{x} \otimes \mathcal{U}_T) \hat{\mathcal{K}}$ since
\begin{equation}
    (\sigma_{x} \otimes \mathcal{U}_T) H_{d-1}(-\mathbf{k}_{\perp})^* = H_{d-1}(\mathbf{k}_{\perp})(\sigma_{x} \otimes \mathcal{U}_T),
\end{equation}
and particle-hole symmetry with $\hat{\mathcal{C}}= (i\sigma_{y} \otimes \mathcal{U}_T) \hat{\mathcal{K}}$ since 
\begin{equation}
    (i\sigma_{y} \otimes \mathcal{U}_T) H_{d-1}(-\mathbf{k}_{\perp})^* = -H_{d-1}(\mathbf{k}_{\perp})(i\sigma_{y} \otimes \mathcal{U}_T).
\end{equation}
Notice that time-reversal and particle-hole symmetries combine into the chiral symmetry as expected. 
The type of time-reversal and particle-hole symmetry of the effective Hamiltonian $H_{d-1}$ can be determined from
\begin{gather}
    (\sigma_{x} \otimes \mathcal{U}_T)^* (\sigma_{x} \otimes \mathcal{U}_T) = \mathbb{I}_{2}\otimes(\mathcal{U}_{T}^* \mathcal{U}_T)  \\
    (i\sigma_{y} \otimes \mathcal{U}_T)^* (i\sigma_{y} \otimes \mathcal{U}_T) = -\mathbb{I}_{2}\otimes(\mathcal{U}_{T}^* \mathcal{U}_T). 
\end{gather}
Hence, dimensional reduction transforms the classes AI and AIII into CI and DIII, respectively.

{\em Classes C and D.---}Next, consider classes C and D with only particle-hole symmetry. Similar to the previous case, particle-hole symmetry implies that the reflection matrix at the Fermi energy obeys 
\begin{gather}
    r(\mathbf{k}_{\perp}) \mathcal{U}_{C}  = \mathcal{U}_{C}r^*(-\mathbf{k}_{\perp})
    \label{c1} \\
    r^{\dagger}(\mathbf{k}_{\perp}) \mathcal{U}_{C}  = \mathcal{U}_{C}r^T(-\mathbf{k}_{\perp}).
\label{c2}
\end{gather}

The effective Hamiltonian $H_{d-1}$ according to Eq.~(\ref{eq:effH_nonchiral}) has chiral symmetry. In addition, we can establish using Eqs.\ (\ref{c1}) and (\ref{c2}) that $H_{d-1}$ has time-reversal symmetry with $\hat{\mathcal{T}}= (\mathbb{I}_{2} \otimes \mathcal{U}_T) \hat{\mathcal{K}}$ since
\begin{equation}
    (\mathbb{I}_{2}\otimes\mathcal{U}_C)H_{d-1}(-\mathbf{k}_{\perp})^* = H_{d-1}(\mathbf{k}_{\perp}) (\mathbb{I}_{2}\otimes\mathcal{U}_C)
\end{equation}
and particle-hole symmetry with $\hat{\mathcal{C}}= (\sigma_{z} \otimes \mathcal{U}_T) \hat{\mathcal{K}}$ since
\begin{equation}
    (\sigma_{z} \otimes \mathcal{U}_C) H_{d-1}(-\mathbf{k}_{\perp})^* = -H_{d-1}(\mathbf{k}_{\perp})(\sigma_{z} \otimes \mathcal{U}_C).
\end{equation}
Time-reversal and particle-hole symmetries combine as  
\begin{equation}
    (\mathbb{I}_{2}\otimes\mathcal{U}_C)(\sigma_{z} \otimes \mathcal{U}_C)^*=\pm \sigma_{z} \otimes \mathbb{I},
\end{equation}
and hence produce the chiral symmetry as expected. 

The types of the new time-reversal and particle-hole symmetry are determined from
\begin{gather}
    (\mathbb{I}_{2}\otimes\mathcal{U}_C)(\mathbb{I}_{2}\otimes\mathcal{U}_C)^* = \mathbb{I}_{2}\otimes (\mathcal{U}_{C}\mathcal{U}_{C}^*) \\
    (\mathbb{\sigma}_{z}\otimes\mathcal{U}_C)(\mathbb{I}_{z}\otimes\mathcal{U}_C)^* = \mathbb{I}_{2}\otimes (\mathcal{U}_{C}\mathcal{U}_{C}^*).
\end{gather}
Hence, classes C and D transform into CII and BDI respectively under the procedure of dimensional reduction.

{\em Classes CI, CII, DIII, and BDI.---}Finally, we focus on the symmetry classes where time-reversal, particle-hole, and chiral symmetries are all present. In this situation, we have $\mathcal{U}_{S}= \mathcal{U}_{T} \mathcal{U}_{C}^*$. Then, $\mathcal{U}_{S}^2=1$ implies $\mathcal{U}_{T}^* \mathcal{U}_{C}\mathcal{U}_{S}^*=1$. This can be used to show that 
\begin{equation}
    \mathcal{U}_{S} \mathcal{U}_{C}  =  \mathcal{U}_{C}\mathcal{U}_S^* 
    (\mathcal{U}_{C}^*\mathcal{U}_{C}) (\mathcal{U}_{T}^*\mathcal{U}_{T}).
    \label{USUSUS}
\end{equation}
Notice that $\mathcal{U}_{C}^*\mathcal{U}_{C}=\pm 1 $ and $\mathcal{U}_{T}^*\mathcal{U}_{T}=\pm 1$ are just numbers. 

The effective Hamiltonian $H_{d-1}$, in this case defined by Eq.~(\ref{eq:effH_chiral}), has no chiral symmetry and can thus have either time-reversal or particle-hole symmetry, but not both. Using Eq.\ (\ref{USUSUS}), we find that $H_{d-1}$ has the property
\begin{equation}
    H_{d-1}(\mathbf{k}_{\perp})\mathcal{U}_{C}  = (\mathcal{U}_{C}^*\mathcal{U}_{C}) (\mathcal{U}_{T}^*\mathcal{U}_{T}) \mathcal{U}_{C} H_{d-1}(-\mathbf{k}_{\perp})^* .
\end{equation}
Thus, $\mathcal{U}_{C}\hat{\mathcal{K}}$ defines a time-reversal symmetry or particle-hole symmetry for $H_{d-1}$, depending on whether the quantity $(\mathcal{U}_{C}^*\mathcal{U}_{C}) (\mathcal{U}_{T}^*\mathcal{U}_{T})$ takes on the value $+1$ or $-1$, respectively. Whether this symmetry squares to $+1$ or $-1$ is determined by $\mathcal{U}_{C}^* \mathcal{U}_C$. This implies that under the procedure of dimensional reduction, classes CI, CII, DIII, and BDI transform into C, AII, D, and AI, respectively.

\begin{table}
\caption{\label{tab:tenfold}Periodic table of topological insulators and superconductors in $d=0,\dots,7$ dimensions. The first column denotes the ten symmetry classes of fermionic Hamiltonians, charactered by the absence ($0$) or  presence ($\pm$ or $1$) of time-reversal ($\hat{\mathcal{T}}$), particle-hole ($\hat{\mathcal{C}}$), and chiral symmetries ($\hat{\mathcal{S}}$). Time reversal and particle hole symmetry exist in two types denoted by $\pm$. Classes which support only trivial phases are denoted by "-",
while classes with nontrivial topological classifications are indicated by the type of topological invariant ($\mathbb{Z}, 2\mathbb{Z}, \mathbb{Z}_2$). }
\begin{ruledtabular}
\centering
\begin{tabular}{c|ccc|cccccccc}
Class & $\hat{\mathcal{T}}$ & $\hat{\mathcal{C}}$ & $\hat{\mathcal{S}}$ & 0 & 1 & 2 & 3 & 4 & 5 & 6 & 7\tabularnewline
\hline 
A & 0 & 0 & 0 & $\mathbb{Z}$ & - & $\mathbb{Z}$ & - & $\mathbb{Z}$ & - & $\mathbb{Z}$ & -\tabularnewline
AIII & 0 & 0 & 1 & - & $\mathbb{Z}$ & - & $\mathbb{Z}$ & - & $\mathbb{Z}$ & - & $\mathbb{Z}$\tabularnewline
\hline 
AI & + & 0 & 0 & $\mathbb{Z}$ & - & - & - & $2\mathbb{Z}$ & - & $\mathbb{Z}_{2}$ & $\mathbb{Z}_{2}$\tabularnewline
BDI & + & + & 1 & $\mathbb{Z}_{2}$ & $\mathbb{Z}$ & - & - & - & $2\mathbb{Z}$ & - & $\mathbb{Z}_{2}$\tabularnewline
D & 0 & + & 0 & $\mathbb{Z}_{2}$ & $\mathbb{Z}_{2}$ & $\mathbb{Z}$ & - & - & - & $2\mathbb{Z}$ & -\tabularnewline
DIII & - & + & 1 & - & $\mathbb{Z}_{2}$ & $\mathbb{Z}_{2}$ & $\mathbb{Z}$ & - & - & - & $2\mathbb{Z}$\tabularnewline
AII & - & 0 & 0 & $2\mathbb{Z}$ & - & $\mathbb{Z}_{2}$ & $\mathbb{Z}_{2}$ & $\mathbb{Z}$ & - & - & -\tabularnewline
CII & - & - & 1 & - & $2\mathbb{Z}$ & - & $\mathbb{Z}_{2}$ & $\mathbb{Z}_{2}$ & $\mathbb{Z}$ & - & -\tabularnewline
C & 0 & - & 0 & - & - & $2\mathbb{Z}$ & - & $\mathbb{Z}_{2}$ & $\mathbb{Z}_{2}$ & $\mathbb{Z}$ & -\tabularnewline
CI & + & - & 1 & - & - & - & $2\mathbb{Z}$ & - & $\mathbb{Z}_{2}$ & $\mathbb{Z}_{2}$ & $\mathbb{Z}$\tabularnewline
\end{tabular}
\end{ruledtabular}
\end{table}

Using this procedure of dimensional reduction, a $d$-dimensional Hamiltonian in one symmetry class is related to a $(d-1)$-dimensional Hamiltonian $H_{d-1}$ in another symmetry class. Yet, both Hamiltonians have the same topological invariants \cite{Fulga2012}. This reproduces the Bott periodicity of the topological classification of symmetry classes \cite{Schnyder2008,Kitaev2009,Ryu2010}. In particular, classes A and AIII are transformed into one another, producing the complex Bott periodicity (mod $2$). The remaining eight classes with antiunitary symmetries are shifted by one in Table  \ref{tab:tenfold}, giving rise to the real Bott periodicity (mod $8$). This is summarized in Table \ref{tab:tenfold}. 

\subsection{Topological invariants}

From 1D Hamiltonians, dimensional reduction produces 0D Hamiltonians. For classes without symmetry between positive and negative eigenenergies (classes A, AI, and AII), the topological invariant in 0D is given by the imbalance between positive and negative eigenenergies (measured relative to the Fermi energy). This  corresponds to a $\mathbb{Z}$ index, except in class AII, where this number is always even because of Kramers degeneracy ($2\mathbb{Z}$ index). Of the remaining classes with symmetry between positive and negative-energy spectra, D and BDI are special as there is no level repulsion between the levels of a pair
with positive and negative energies \cite{Altland97,Beenakker2015}. These classes have a $\mathbb{Z}_2$ topological invariant, which is associated with fermion parity and can be expressed as the sign of a Pfaffian
\begin{equation}
    \mathcal{Q} =  \Sgn[\mathrm{Pf}(iH_{0})],
    \label{eq:pfaffian}
\end{equation}
where the Hamiltonian $H_{0}$ is the zero-dimensional Hamiltonian, which is antisymmetric in an appropriate basis. The remaining five symmetry classes are always topologically trivial. 

\subsubsection{1D systems}

The dimensional reduction implies that in one dimension, there are topological phases in symmetry classes AIII, BDI, CII, DIII, and D. The topological invariants in 0D immediately provide explicit expressions for boundary topological invariants in 1D.

Dimensional reduction connects the classes AIII, BDI, and CII in 1D to classes with $\mathbb{Z}$ or $2\mathbb{Z}$ indices in 0D. Using Eq.\ (\ref{eq:effH_chiral}), the 1D classes are therefore characterized by the boundary topological invariant 
\begin{equation}
    \mathcal{Q} = \nu(\lim_{\omega\to 0} \mathcal{U}_S \frac{\mathbb{I}-iVGV^{\dagger}}{\mathbb{I}+iVGV^\dagger}) - \frac{M}{2},
    \label{eq:Q_BDI_1D}
\end{equation}
where $\nu(A)$ denotes the number of negative eigenvalues of the hermitian matrix $A$. The constant $M/2$ ensures that the trivial phase has topological invariant $\mathcal{Q}=0$. This invariant indeed counts the imbalance between positive and negative eigenenergies of $H_{d-1}$.

By the bulk-boundary correspondence, these topological invariants can also be computed from the bulk 1D Hamiltonian $H(k)$. Due to the presence of chiral symmetry, the bulk Hamiltonian can be written as
\begin{equation}
    H(k)  = \left(
    \begin{array}{cc}
        0   &h(k) \\
        h^{\dagger}(k) &0 
    \end{array}\right),
    \label{eq:H_chiral}
\end{equation}
and the corresponding bulk topological invariant is given by \cite{Zak1989,Ryu2002}
\begin{equation}
    \mathcal{Q} = \frac{1}{2\pi i}\int_{0}^{2\pi} dk\,\frac{d}{dk}\log \det h(k),
    \label{eq:1Dbulk_winding}
\end{equation}
which is a winding number.

In class DIII, $H_{d-1}$ has particle-hole symmetry realized via a unitary matrix $\mathcal{U}_{C}$ with $\mathcal{U}_{C}^* \mathcal{U}_C = 1$. This implies that $\mathcal{U}_C = \mathcal{U}_C^T$, so that it is possible to write $\mathcal{U}_{C} = \mathcal{V}_C \mathcal{V}_C^T$ with a unitary matrix $\mathcal{V}_C$. By the unitary transformation 
\begin{equation*}
H_{d-1} \to \mathcal{V}_C^\dagger H_{d-1} \mathcal{V}_C,
\end{equation*}
$H_{d-1}$ becomes antisymmetric and purely imaginary. According to Eq.~(\ref{eq:pfaffian}), the boundary  topological invariant for class DIII can thus be expressed as 
\begin{equation}
    \mathcal{Q} =  \mathrm{Pf}(i H_{d-1}) =\mathrm{Pf}(\lim_{\omega \to 0}i\mathcal{V}_C^{\dagger} \mathcal{U}_{S} \frac{\mathbb{I}-iVGV^{\dagger}}{\mathbb{I}+iVGV^\dagger}
    \mathcal{V}_C).
\end{equation}
Again, this invariant can also be computed from the bulk Hamiltonian $H(k)$ as \cite{Qi2010}
\begin{equation}
    \mathcal{Q}= \frac{\mathrm{Pf}(\mathcal{U}_T h(\pi))}{\mathrm{Pf}(\mathcal{U}_T h(0))}
    \frac{\sqrt{\det h(0)}}{\sqrt{\det h(\pi)}},
    \label{eq:H_DIII_1D}
\end{equation}
where $h(k)$ is defined as in Eq.~(\ref{eq:H_chiral}).

Class D also allows us to transform $H_{d-1}$ as 
\begin{align}
    H_{d-1} &\to 
    \left(\begin{array}{cc}
        \mathcal{V}_C &0\\
       0 &i\mathcal{V}_C
   \end{array}\right)^{\dagger}
   H_{d-1}
    \left(\begin{array}{cc}
        \mathcal{V}_C &0\\
       0 &i\mathcal{V}_C
   \end{array}\right) \nonumber \\
   &= \left(\begin{array}{cc}
       0 &i\mathcal{V}_C^\dagger r \mathcal{V}_C\\
       -i\mathcal{V}_C^\dagger r^{\dagger} \mathcal{V}_C &0
   \end{array}\right) , 
\end{align}
where $\mathcal{V}_C^\dagger r \mathcal{V}_C$ is purely real. Thus, the boundary topological invariant takes the form
\begin{equation}
    \mathcal{Q} = \mathrm{Pf}(i H_{d-1})= \det\left( \lim_{\omega \to 0}\frac{\mathbb{I}-iVGV^{\dagger}}{\mathbb{I}+iVGV^\dagger}\right)
    \label{eq:1D_clsD}
\end{equation}
according to Eq.~(\ref{eq:pfaffian}). The corresponding bulk topological invariant is \cite{Kitaev2001}
\begin{equation}
    \mathcal{Q} = \Sgn \left(\frac{\mathrm{Pf}H(0)}{\mathrm{Pf}H(\pi)}\right).
\end{equation}

\subsubsection{2D systems}

By the procedure of dimensional reduction, the boundary invariants of 2D topological phases correspond to bulk invariants of 1D Hamiltonians. Using the 1D bulk invariant for classes AIII, BDI, and CII in Eq.~(\ref{eq:1Dbulk_winding}), we obtain the boundary topological invariants for 2D systems in classes A, C, and D expressed in terms of the boundary Green function,
\begin{equation}
    \mathcal{Q} = \frac{1}{2\pi i} \int _{0}^{2\pi}dk\,\frac{d}{dk} \Tr \log \left(\lim_{\omega\to 0} \frac{\mathbb{I}-iVGV^{\dagger}}{\mathbb{I}+iVGV^\dagger}\right).
    \label{eq:Q_CI}
\end{equation}
These boundary topological invariants can again alternatively be computed from a bulk topological index expressed in terms of the bulk Hamiltonian. This bulk topological index takes the form of a Chern number \cite{Chiu2016},
\begin{equation}
    \mathcal{Q} = \frac{1}{2\pi}\int dk_{x} dk_{y} F_{k_x,k_{y}}
\end{equation}
where the Berry curvature is defined as
\begin{equation}
    F_{k_x,k_{y}}  = \sum_{\alpha \in \mathrm{occ.}} i\partial_{k_{x}}\bra{u_\alpha(\v{k})}\partial_{k_{y}}\ket{u_\alpha(\v{k})} - (k_{x}\leftrightarrow k_{y}).
\end{equation}
Here, $\ket{u_{\alpha}(\v{k})}$ denotes the Bloch wave function of the $\alpha$ band. Note that in computing the Berry curvature, one sums over occupied bands only. 

For 2D system in class DIII, the boundary topological invariant follows from the 1D bulk invariant for class D in Eq.~(\ref{eq:1D_clsD}). This yields
\begin{equation}
    \mathcal{Q} =  \Sgn\left(\lim_{\omega \to 0} \frac{\mathrm{Pf}(\mathcal{V}_C^{\dagger} \mathcal{U}_{S} \frac{\mathbb{I}-iVG(0)V^{\dagger}}{\mathbb{I}+iVG(0)V^\dagger}
    \mathcal{V}_C)}{\mathrm{Pf}(\mathcal{V}_C^{\dagger} \mathcal{U}_{S} \frac{\mathbb{I}-iVG(\pi)V^{\dagger}}{\mathbb{I}+iVG(\pi)V^\dagger}
    \mathcal{V}_C)}\right).
\end{equation}
The corresponding bulk topological invariant of the bulk Hamiltonian $H(k_1,k_2)$ can be written as a product of topological invariants of 1D Hamiltonians \cite{Qi2010}
\begin{equation}
    \mathcal{Q}(H(k_1,k_2)) = \mathcal{Q}(H(k_1,0))\mathcal{Q}(H(k_1,\pi)),
\end{equation}
where $\mathcal{Q}(H(k_1,k_2))$ with $k_2=0,\pi$  is given in Eq.~(\ref{eq:H_DIII_1D}).

Finally,  the topological invariant of 2D system in class AII is given by
\begin{equation}
    \mathcal{Q} = \lim_{\omega \to 0}
     \frac{\mathrm{Pf}(\mathcal{U}_T \frac{\mathbb{I}-iVG(\pi)V^{\dagger}}{\mathbb{I}+iVG(\pi)V^\dagger})}{\mathrm{Pf}(\mathcal{U}_T\frac{\mathbb{I}-iVG(0)V^{\dagger}}{\mathbb{I}+iVG(0)V^\dagger})}
     \frac{\sqrt{\det \frac{\mathbb{I}-iVG(0)V^{\dagger}}{\mathbb{I}+iVG(0)V^\dagger}}}{\sqrt{\det \frac{\mathbb{I}-iVG(\pi)V^{\dagger}}{\mathbb{I}+iVG(\pi)V^\dagger}}},
\end{equation}
according to the 1D bulk topological invariant for class DIII in Eq.~(\ref{eq:H_DIII_1D}). This topological invariant can be computed from the bulk via \cite{Fu2006}
\begin{equation}
    \mathcal{Q} = \prod_{\mathbf{K}}\frac{\mathrm{Pf}(w(\v{K}))}{\sqrt{\det(w(\v{K}))}},
\end{equation}
where the matrix element of $w$ is defined as
\begin{equation}
    w_{mn}(\v{k})  = \bra{u_{m}(-\v{k})}\hat{\mathcal{T}}\ket{u_{n}(\v{k})},
\end{equation}
with occupied Bloch bands $\ket{u_n(\v{k})}$.

\subsubsection{3D systems}

The 3D topological invariant in class AII, which transforms into DIII under dimensional reduction, takes the form
\begin{equation}
  \mathcal{Q} = \mathcal{Q}(0) \mathcal{Q}(\pi)
\end{equation}
with
\begin{equation}
  \mathcal{Q}(k) =  
\lim_{\omega \to 0}
     \frac{\mathrm{Pf}(\mathcal{U}_T
     \frac{\mathbb{I}-iVG(\pi,k)V^{\dagger}}{\mathbb{I}+iVG(\pi,k)V^\dagger})}{\mathrm{Pf}(\mathcal{U}_T\frac{\mathbb{I}-iVG(0,k)V^{\dagger}}{\mathbb{I}+iVG(0,k)V^\dagger})}
     \frac{\sqrt{\det \frac{\mathbb{I}-iVG(0,k)V^{\dagger}}{\mathbb{I}+iVG(0,k)V^\dagger}}}{\sqrt{\det
       \frac{\mathbb{I}-iVG(\pi,k)V^{\dagger}}{\mathbb{I}+iVG(\pi,k)V^\dagger}}}.
\end{equation}
The topological invariants in classes AIII, CI, DIII, and CII can be written in terms of 2D bulk topological invariants. For simplicity, we do not write them down explictly here.

\section{Condition for a vanishing bulk gap \label{app:Lyapunov}}

In this appendix, we show that when the bulk band gap vanishes, the matrix $T$ defined in Eq.~(\ref{eq:M_homogeneous}) has at least one eigenvalue with unit modulus and thus one vanishing Lyapunov exponent. This implies that the topological phase transition with its associated gap closing is signalled by at least one vanishing Lyapunov exponent.

Consider a homogeneous, quasi-1D system with periodic boundary conditions. According to the Bloch theorem, we can write the wave function as as $\psi(n) = e^{ikn} \mathbf{u}$, where $\mathbf{u}$ is an $M$-component column vector. Using Eq.~(\ref{eq:SEQ}), we find a zero-energy eigenstate when 
\begin{equation}
    \left[g^{-1}(\omega=0) - Ve^{-ik} -V^{\dagger}e^{ik}\right] \mathbf{u}=0.
\end{equation}
This implies
\begin{equation}
    \det\left[g^{-1}(\omega = 0) - Ve^{-ik} -V^{\dagger}e^{ik}\right] =0.
\end{equation}
Thus, $\lambda = e^{-ik}$ is an eigenvalue of $T(\omega = 0)$ defined in Eq.~(\ref{eq:M_homogeneous}), which implies that the corresponding Lyapunov exponent vanishes.

\section{Derivation of Eq.~(\ref{eq:Ninf})\label{app:proveM_U}}

From Eq.~(\ref{eq:MT}) and (\ref{eq:eigenT}),
we have
\begin{equation}
  \mathcal{M}_{aN}\left(\begin{array}{cc}
    U_{11} & U_{12}\\
    U_{21} & U_{22}
    \end{array}\right)=
    \left(\begin{array}{cc}
        U_{11} & U_{12}\\
        U_{21} & U_{22}
        \end{array}\right)
    \left(\begin{array}{cc}
      \Lambda^{N} & 0\\
        0 & (\Lambda^{-N})^{*}
      \end{array}\right)
  \end{equation}
which gives
\begin{gather}
  {\cal M}_{aN,11}=U_{11}\Lambda^{N}(U^{-1})_{11}+U_{12}(\Lambda^{*})^{-N}(U^{-1})_{21}\\
  {\cal
  M}_{aN,21}=U_{21}\Lambda^{N}(U^{-1})_{11}+U_{22}(\Lambda^{*})^{-N}(U^{-1})_{21}.
\end{gather}
Hence
\begin{widetext}
  \begin{align}
    \mathcal{M}_{aN,21}\mathcal{M}_{aN,11}^{-1}&=\left[U_{21}\Lambda^{N}(U^{-1})_{11}+U_{22}(\Lambda^{*})^{-N}(U^{-1})_{21}\right]\left[U_{11}\Lambda^{N}(U^{-1})_{11}+U_{12}(\Lambda^{*})^{-    N}(U^{-1})_{21}\right]^{-1}
    \nonumber \\
    &=\left[U_{21}\Lambda^{N}(U^{-1})_{11}+O(\Lambda^{-N})\right]\left[(U^{-1})_{11}^{-1}\Lambda^{-N}U_{11}^{-1}+O(\Lambda^{-2N})\right]
    \nonumber\\
    &=U_{21}U_{11}^{-1}+O((\Lambda^{-2N}).
  \end{align}
\end{widetext}
Taking the thermodynamic limit $N\to \infty$, we have $\Lambda^{-2N}\to 0$ and thus
arrive at Eq.~(\ref{eq:Ninf}).

\bibliographystyle{apsrev4-1}

\begin{thebibliography}{35}%
\makeatletter
\providecommand \@ifxundefined [1]{%
 \@ifx{#1\undefined}
}%
\providecommand \@ifnum [1]{%
 \ifnum #1\expandafter \@firstoftwo
 \else \expandafter \@secondoftwo
 \fi
}%
\providecommand \@ifx [1]{%
 \ifx #1\expandafter \@firstoftwo
 \else \expandafter \@secondoftwo
 \fi
}%
\providecommand \natexlab [1]{#1}%
\providecommand \enquote  [1]{``#1''}%
\providecommand \bibnamefont  [1]{#1}%
\providecommand \bibfnamefont [1]{#1}%
\providecommand \citenamefont [1]{#1}%
\providecommand \href@noop [0]{\@secondoftwo}%
\providecommand \href [0]{\begingroup \@sanitize@url \@href}%
\providecommand \@href[1]{\@@startlink{#1}\@@href}%
\providecommand \@@href[1]{\endgroup#1\@@endlink}%
\providecommand \@sanitize@url [0]{\catcode `\\12\catcode `\$12\catcode
  `\&12\catcode `\#12\catcode `\^12\catcode `\_12\catcode `\%12\relax}%
\providecommand \@@startlink[1]{}%
\providecommand \@@endlink[0]{}%
\providecommand \url  [0]{\begingroup\@sanitize@url \@url }%
\providecommand \@url [1]{\endgroup\@href {#1}{\urlprefix }}%
\providecommand \urlprefix  [0]{URL }%
\providecommand \Eprint [0]{\href }%
\providecommand \doibase [0]{http://dx.doi.org/}%
\providecommand \selectlanguage [0]{\@gobble}%
\providecommand \bibinfo  [0]{\@secondoftwo}%
\providecommand \bibfield  [0]{\@secondoftwo}%
\providecommand \translation [1]{[#1]}%
\providecommand \BibitemOpen [0]{}%
\providecommand \bibitemStop [0]{}%
\providecommand \bibitemNoStop [0]{.\EOS\space}%
\providecommand \EOS [0]{\spacefactor3000\relax}%
\providecommand \BibitemShut  [1]{\csname bibitem#1\endcsname}%
\let\auto@bib@innerbib\@empty
\bibitem [{\citenamefont {Hasan}\ and\ \citenamefont {Kane}(2010)}]{Hasan2010}%
  \BibitemOpen
  \bibfield  {author} {\bibinfo {author} {\bibfnamefont {M.~Z.}\ \bibnamefont
  {Hasan}}\ and\ \bibinfo {author} {\bibfnamefont {C.~L.}\ \bibnamefont
  {Kane}},\ }\href {\doibase 10.1103/RevModPhys.82.3045} {\bibfield  {journal}
  {\bibinfo  {journal} {Rev. Mod. Phys.}\ }\textbf {\bibinfo {volume} {82}},\
  \bibinfo {pages} {3045} (\bibinfo {year} {2010})}\BibitemShut {NoStop}%
\bibitem [{\citenamefont {Qi}\ and\ \citenamefont {Zhang}(2011)}]{Qi2011}%
  \BibitemOpen
  \bibfield  {author} {\bibinfo {author} {\bibfnamefont {X.-L.}\ \bibnamefont
  {Qi}}\ and\ \bibinfo {author} {\bibfnamefont {S.-C.}\ \bibnamefont {Zhang}},\
  }\href {\doibase 10.1103/RevModPhys.83.1057} {\bibfield  {journal} {\bibinfo
  {journal} {Rev. Mod. Phys.}\ }\textbf {\bibinfo {volume} {83}},\ \bibinfo
  {pages} {1057} (\bibinfo {year} {2011})}\BibitemShut {NoStop}%
\bibitem [{\citenamefont {Bernevig}\ and\ \citenamefont
  {Hughes}(2013)}]{Bernevig2013book}%
  \BibitemOpen
  \bibfield  {author} {\bibinfo {author} {\bibfnamefont {B.~A.}\ \bibnamefont
  {Bernevig}}\ and\ \bibinfo {author} {\bibfnamefont {T.~L.}\ \bibnamefont
  {Hughes}},\ }\href@noop {} {\emph {\bibinfo {title} {Topological insulators
  and topological superconductors}}}\ (\bibinfo  {publisher} {Princeton
  University Press},\ \bibinfo {year} {2013})\BibitemShut {NoStop}%
\bibitem [{\citenamefont {Schnyder}\ \emph {et~al.}(2008)\citenamefont
  {Schnyder}, \citenamefont {Ryu}, \citenamefont {Furusaki},\ and\
  \citenamefont {Ludwig}}]{Schnyder2008}%
  \BibitemOpen
  \bibfield  {author} {\bibinfo {author} {\bibfnamefont {A.~P.}\ \bibnamefont
  {Schnyder}}, \bibinfo {author} {\bibfnamefont {S.}~\bibnamefont {Ryu}},
  \bibinfo {author} {\bibfnamefont {A.}~\bibnamefont {Furusaki}}, \ and\
  \bibinfo {author} {\bibfnamefont {A.~W.~W.}\ \bibnamefont {Ludwig}},\ }\href
  {\doibase 10.1103/PhysRevB.78.195125} {\bibfield  {journal} {\bibinfo
  {journal} {Phys. Rev. B}\ }\textbf {\bibinfo {volume} {78}},\ \bibinfo
  {pages} {195125} (\bibinfo {year} {2008})}\BibitemShut {NoStop}%
\bibitem [{\citenamefont {Kitaev}(2009)}]{Kitaev2009}%
  \BibitemOpen
  \bibfield  {author} {\bibinfo {author} {\bibfnamefont {A.}~\bibnamefont
  {Kitaev}},\ }\href@noop {} {\bibfield  {journal} {\bibinfo  {journal} {AIP
  Conf. Proc.}\ }\textbf {\bibinfo {volume} {1134}},\ \bibinfo {pages} {22}
  (\bibinfo {year} {2009})}\BibitemShut {NoStop}%
\bibitem [{\citenamefont {Ryu}\ \emph {et~al.}(2010)\citenamefont {Ryu},
  \citenamefont {Schnyder}, \citenamefont {Furusaki},\ and\ \citenamefont
  {Ludwig}}]{Ryu2010}%
  \BibitemOpen
  \bibfield  {author} {\bibinfo {author} {\bibfnamefont {S.}~\bibnamefont
  {Ryu}}, \bibinfo {author} {\bibfnamefont {A.~P.}\ \bibnamefont {Schnyder}},
  \bibinfo {author} {\bibfnamefont {A.}~\bibnamefont {Furusaki}}, \ and\
  \bibinfo {author} {\bibfnamefont {A.~W.}\ \bibnamefont {Ludwig}},\
  }\href@noop {} {\bibfield  {journal} {\bibinfo  {journal} {New Journal of
  Physics}\ }\textbf {\bibinfo {volume} {12}},\ \bibinfo {pages} {065010}
  (\bibinfo {year} {2010})}\BibitemShut {NoStop}%
\bibitem [{\citenamefont {Teo}\ and\ \citenamefont {Kane}(2010)}]{Teo2010}%
  \BibitemOpen
  \bibfield  {author} {\bibinfo {author} {\bibfnamefont {J.~C.~Y.}\
  \bibnamefont {Teo}}\ and\ \bibinfo {author} {\bibfnamefont {C.~L.}\
  \bibnamefont {Kane}},\ }\href {\doibase 10.1103/PhysRevB.82.115120}
  {\bibfield  {journal} {\bibinfo  {journal} {Phys. Rev. B}\ }\textbf {\bibinfo
  {volume} {82}},\ \bibinfo {pages} {115120} (\bibinfo {year}
  {2010})}\BibitemShut {NoStop}%
\bibitem [{\citenamefont {Essin}\ and\ \citenamefont
  {Gurarie}(2011)}]{Essin2011}%
  \BibitemOpen
  \bibfield  {author} {\bibinfo {author} {\bibfnamefont {A.~M.}\ \bibnamefont
  {Essin}}\ and\ \bibinfo {author} {\bibfnamefont {V.}~\bibnamefont
  {Gurarie}},\ }\href {\doibase 10.1103/PhysRevB.84.125132} {\bibfield
  {journal} {\bibinfo  {journal} {Phys. Rev. B}\ }\textbf {\bibinfo {volume}
  {84}},\ \bibinfo {pages} {125132} (\bibinfo {year} {2011})}\BibitemShut
  {NoStop}%
\bibitem [{\citenamefont {Fulga}\ \emph {et~al.}(2012)\citenamefont {Fulga},
  \citenamefont {Hassler},\ and\ \citenamefont {Akhmerov}}]{Fulga2012}%
  \BibitemOpen
  \bibfield  {author} {\bibinfo {author} {\bibfnamefont {I.~C.}\ \bibnamefont
  {Fulga}}, \bibinfo {author} {\bibfnamefont {F.}~\bibnamefont {Hassler}}, \
  and\ \bibinfo {author} {\bibfnamefont {A.~R.}\ \bibnamefont {Akhmerov}},\
  }\href {\doibase 10.1103/PhysRevB.85.165409} {\bibfield  {journal} {\bibinfo
  {journal} {Phys. Rev. B}\ }\textbf {\bibinfo {volume} {85}},\ \bibinfo
  {pages} {165409} (\bibinfo {year} {2012})}\BibitemShut {NoStop}%
\bibitem [{\citenamefont {Chiu}\ \emph {et~al.}(2016)\citenamefont {Chiu},
  \citenamefont {Teo}, \citenamefont {Schnyder},\ and\ \citenamefont
  {Ryu}}]{Chiu2016}%
  \BibitemOpen
  \bibfield  {author} {\bibinfo {author} {\bibfnamefont {C.-K.}\ \bibnamefont
  {Chiu}}, \bibinfo {author} {\bibfnamefont {J.~C.~Y.}\ \bibnamefont {Teo}},
  \bibinfo {author} {\bibfnamefont {A.~P.}\ \bibnamefont {Schnyder}}, \ and\
  \bibinfo {author} {\bibfnamefont {S.}~\bibnamefont {Ryu}},\ }\href {\doibase
  10.1103/RevModPhys.88.035005} {\bibfield  {journal} {\bibinfo  {journal}
  {Rev. Mod. Phys.}\ }\textbf {\bibinfo {volume} {88}},\ \bibinfo {pages}
  {035005} (\bibinfo {year} {2016})}\BibitemShut {NoStop}%
\bibitem [{\citenamefont {Sancho}\ \emph {et~al.}(1985)\citenamefont {Sancho},
  \citenamefont {Sancho}, \citenamefont {Sancho},\ and\ \citenamefont
  {Rubio}}]{Sancho1985}%
  \BibitemOpen
  \bibfield  {author} {\bibinfo {author} {\bibfnamefont {M.~L.}\ \bibnamefont
  {Sancho}}, \bibinfo {author} {\bibfnamefont {J.~L.}\ \bibnamefont {Sancho}},
  \bibinfo {author} {\bibfnamefont {J.~L.}\ \bibnamefont {Sancho}}, \ and\
  \bibinfo {author} {\bibfnamefont {J.}~\bibnamefont {Rubio}},\ }\href@noop {}
  {\bibfield  {journal} {\bibinfo  {journal} {J. Phys. F: Met. Phys.}\ }\textbf
  {\bibinfo {volume} {15}},\ \bibinfo {pages} {851} (\bibinfo {year}
  {1985})}\BibitemShut {NoStop}%
\bibitem [{\citenamefont {Umerski}(1997)}]{Umerski1997}%
  \BibitemOpen
  \bibfield  {author} {\bibinfo {author} {\bibfnamefont {A.}~\bibnamefont
  {Umerski}},\ }\href {\doibase 10.1103/PhysRevB.55.5266} {\bibfield  {journal}
  {\bibinfo  {journal} {Phys. Rev. B}\ }\textbf {\bibinfo {volume} {55}},\
  \bibinfo {pages} {5266} (\bibinfo {year} {1997})}\BibitemShut {NoStop}%
\bibitem [{\citenamefont {Su}\ \emph {et~al.}(1979)\citenamefont {Su},
  \citenamefont {Schrieffer},\ and\ \citenamefont {Heeger}}]{Su1979}%
  \BibitemOpen
  \bibfield  {author} {\bibinfo {author} {\bibfnamefont {W.~P.}\ \bibnamefont
  {Su}}, \bibinfo {author} {\bibfnamefont {J.~R.}\ \bibnamefont {Schrieffer}},
  \ and\ \bibinfo {author} {\bibfnamefont {A.~J.}\ \bibnamefont {Heeger}},\
  }\href {\doibase 10.1103/PhysRevLett.42.1698} {\bibfield  {journal} {\bibinfo
   {journal} {Phys. Rev. Lett.}\ }\textbf {\bibinfo {volume} {42}},\ \bibinfo
  {pages} {1698} (\bibinfo {year} {1979})}\BibitemShut {NoStop}%
\bibitem [{\citenamefont {Pichard}\ and\ \citenamefont
  {Sarma}(1981)}]{Pichard1981}%
  \BibitemOpen
  \bibfield  {author} {\bibinfo {author} {\bibfnamefont {J.}~\bibnamefont
  {Pichard}}\ and\ \bibinfo {author} {\bibfnamefont {G.}~\bibnamefont
  {Sarma}},\ }\href@noop {} {\bibfield  {journal} {\bibinfo  {journal} {Journal
  of Physics C: Solid State Physics}\ }\textbf {\bibinfo {volume} {14}},\
  \bibinfo {pages} {L127} (\bibinfo {year} {1981})}\BibitemShut {NoStop}%
\bibitem [{\citenamefont {Lee}\ and\ \citenamefont
  {Joannopoulos}(1981)}]{Lee1981}%
  \BibitemOpen
  \bibfield  {author} {\bibinfo {author} {\bibfnamefont {D.~H.}\ \bibnamefont
  {Lee}}\ and\ \bibinfo {author} {\bibfnamefont {J.~D.}\ \bibnamefont
  {Joannopoulos}},\ }\href {\doibase 10.1103/PhysRevB.23.4988} {\bibfield
  {journal} {\bibinfo  {journal} {Phys. Rev. B}\ }\textbf {\bibinfo {volume}
  {23}},\ \bibinfo {pages} {4988} (\bibinfo {year} {1981})}\BibitemShut
  {NoStop}%
\bibitem [{\citenamefont {Hatsugai}(1993{\natexlab{a}})}]{Hatsugai1993a}%
  \BibitemOpen
  \bibfield  {author} {\bibinfo {author} {\bibfnamefont {Y.}~\bibnamefont
  {Hatsugai}},\ }\href {\doibase 10.1103/PhysRevLett.71.3697} {\bibfield
  {journal} {\bibinfo  {journal} {Phys. Rev. Lett.}\ }\textbf {\bibinfo
  {volume} {71}},\ \bibinfo {pages} {3697} (\bibinfo {year}
  {1993}{\natexlab{a}})}\BibitemShut {NoStop}%
\bibitem [{\citenamefont {Hatsugai}(1993{\natexlab{b}})}]{Hatsugai1993b}%
  \BibitemOpen
  \bibfield  {author} {\bibinfo {author} {\bibfnamefont {Y.}~\bibnamefont
  {Hatsugai}},\ }\href {\doibase 10.1103/PhysRevB.48.11851} {\bibfield
  {journal} {\bibinfo  {journal} {Phys. Rev. B}\ }\textbf {\bibinfo {volume}
  {48}},\ \bibinfo {pages} {11851} (\bibinfo {year}
  {1993}{\natexlab{b}})}\BibitemShut {NoStop}%
\bibitem [{\citenamefont {Kitaev}(2001)}]{Kitaev2001}%
  \BibitemOpen
  \bibfield  {author} {\bibinfo {author} {\bibfnamefont {A.}~\bibnamefont
  {Kitaev}},\ }\href {http://stacks.iop.org/1063-7869/44/i=10S/a=S29}
  {\bibfield  {journal} {\bibinfo  {journal} {Phys. Usp.}\ }\textbf {\bibinfo
  {volume} {44}},\ \bibinfo {pages} {131} (\bibinfo {year} {2001})}\BibitemShut
  {NoStop}%
\bibitem [{\citenamefont {Benalcazar}\ \emph {et~al.}(2016)\citenamefont
  {Benalcazar}, \citenamefont {Bernevig},\ and\ \citenamefont
  {Hughes}}]{Benalcazar2016}%
  \BibitemOpen
  \bibfield  {author} {\bibinfo {author} {\bibfnamefont {W.~A.}\ \bibnamefont
  {Benalcazar}}, \bibinfo {author} {\bibfnamefont {B.~A.}\ \bibnamefont
  {Bernevig}}, \ and\ \bibinfo {author} {\bibfnamefont {T.~L.}\ \bibnamefont
  {Hughes}},\ }\href@noop {} {\bibfield  {journal} {\bibinfo  {journal}
  {arXiv:1611.07987}\ } (\bibinfo {year} {2016})}\BibitemShut {NoStop}%
\bibitem [{\citenamefont {Dai}\ \emph {et~al.}(2008)\citenamefont {Dai},
  \citenamefont {Hughes}, \citenamefont {Qi}, \citenamefont {Fang},\ and\
  \citenamefont {Zhang}}]{Dai2008}%
  \BibitemOpen
  \bibfield  {author} {\bibinfo {author} {\bibfnamefont {X.}~\bibnamefont
  {Dai}}, \bibinfo {author} {\bibfnamefont {T.~L.}\ \bibnamefont {Hughes}},
  \bibinfo {author} {\bibfnamefont {X.-L.}\ \bibnamefont {Qi}}, \bibinfo
  {author} {\bibfnamefont {Z.}~\bibnamefont {Fang}}, \ and\ \bibinfo {author}
  {\bibfnamefont {S.-C.}\ \bibnamefont {Zhang}},\ }\href {\doibase
  10.1103/PhysRevB.77.125319} {\bibfield  {journal} {\bibinfo  {journal} {Phys.
  Rev. B}\ }\textbf {\bibinfo {volume} {77}},\ \bibinfo {pages} {125319}
  (\bibinfo {year} {2008})}\BibitemShut {NoStop}%
\bibitem [{\citenamefont {Kim}\ \emph {et~al.}(2015)\citenamefont {Kim},
  \citenamefont {Mong}, \citenamefont {Franz},\ and\ \citenamefont
  {Refael}}]{Kim2015}%
  \BibitemOpen
  \bibfield  {author} {\bibinfo {author} {\bibfnamefont {K.~W.}\ \bibnamefont
  {Kim}}, \bibinfo {author} {\bibfnamefont {R.~S.~K.}\ \bibnamefont {Mong}},
  \bibinfo {author} {\bibfnamefont {M.}~\bibnamefont {Franz}}, \ and\ \bibinfo
  {author} {\bibfnamefont {G.}~\bibnamefont {Refael}},\ }\href {\doibase
  10.1103/PhysRevB.92.075110} {\bibfield  {journal} {\bibinfo  {journal} {Phys.
  Rev. B}\ }\textbf {\bibinfo {volume} {92}},\ \bibinfo {pages} {075110}
  (\bibinfo {year} {2015})}\BibitemShut {NoStop}%
\bibitem [{\citenamefont {Mahaux}\ and\ \citenamefont
  {Weidenm{\"u}ller}(1969)}]{Mahaux1969}%
  \BibitemOpen
  \bibfield  {author} {\bibinfo {author} {\bibfnamefont {C.}~\bibnamefont
  {Mahaux}}\ and\ \bibinfo {author} {\bibfnamefont {H.}~\bibnamefont
  {Weidenm{\"u}ller}},\ }\href {https://books.google.de/books?id=PtW6AAAAIAAJ}
  {\emph {\bibinfo {title} {Shell-model approach to nuclear reactions}}}\
  (\bibinfo  {publisher} {North-Holland Pub. Co.},\ \bibinfo {year}
  {1969})\BibitemShut {NoStop}%
\bibitem [{\citenamefont {Aleiner}\ \emph {et~al.}(2002)\citenamefont
  {Aleiner}, \citenamefont {Brouwer},\ and\ \citenamefont
  {Glazman}}]{Aleiner2002}%
  \BibitemOpen
  \bibfield  {author} {\bibinfo {author} {\bibfnamefont {I.~L.}\ \bibnamefont
  {Aleiner}}, \bibinfo {author} {\bibfnamefont {P.~W.}\ \bibnamefont
  {Brouwer}}, \ and\ \bibinfo {author} {\bibfnamefont {L.~I.}\ \bibnamefont
  {Glazman}},\ }\href@noop {} {\bibfield  {journal} {\bibinfo  {journal} {Phys.
  Rep.}\ }\textbf {\bibinfo {volume} {358}},\ \bibinfo {pages} {309} (\bibinfo
  {year} {2002})}\BibitemShut {NoStop}%
\bibitem [{\citenamefont {Boffi}\ \emph {et~al.}(2014)\citenamefont {Boffi},
  \citenamefont {Hill},\ and\ \citenamefont {Reuter}}]{Boffi2014}%
  \BibitemOpen
  \bibfield  {author} {\bibinfo {author} {\bibfnamefont {N.~M.}\ \bibnamefont
  {Boffi}}, \bibinfo {author} {\bibfnamefont {J.~C.}\ \bibnamefont {Hill}}, \
  and\ \bibinfo {author} {\bibfnamefont {M.~G.}\ \bibnamefont {Reuter}},\
  }\href@noop {} {\bibfield  {journal} {\bibinfo  {journal} {Comput. Sci.
  Discov.}\ }\textbf {\bibinfo {volume} {8}},\ \bibinfo {pages} {015001}
  (\bibinfo {year} {2014})}\BibitemShut {NoStop}%
\bibitem [{\citenamefont {Beenakker}(1997)}]{Beenakker1997}%
  \BibitemOpen
  \bibfield  {author} {\bibinfo {author} {\bibfnamefont {C.~W.~J.}\
  \bibnamefont {Beenakker}},\ }\href {\doibase 10.1103/RevModPhys.69.731}
  {\bibfield  {journal} {\bibinfo  {journal} {Rev. Mod. Phys.}\ }\textbf
  {\bibinfo {volume} {69}},\ \bibinfo {pages} {731} (\bibinfo {year}
  {1997})}\BibitemShut {NoStop}%
\bibitem [{\citenamefont {Zazunov}\ \emph {et~al.}(2016)\citenamefont
  {Zazunov}, \citenamefont {Egger},\ and\ \citenamefont
  {Levy~Yeyati}}]{Zazunov2016}%
  \BibitemOpen
  \bibfield  {author} {\bibinfo {author} {\bibfnamefont {A.}~\bibnamefont
  {Zazunov}}, \bibinfo {author} {\bibfnamefont {R.}~\bibnamefont {Egger}}, \
  and\ \bibinfo {author} {\bibfnamefont {A.}~\bibnamefont {Levy~Yeyati}},\
  }\href {\doibase 10.1103/PhysRevB.94.014502} {\bibfield  {journal} {\bibinfo
  {journal} {Phys. Rev. B}\ }\textbf {\bibinfo {volume} {94}},\ \bibinfo
  {pages} {014502} (\bibinfo {year} {2016})}\BibitemShut {NoStop}%
\bibitem [{\citenamefont {Laughlin}(1981)}]{Laughlin1981}%
  \BibitemOpen
  \bibfield  {author} {\bibinfo {author} {\bibfnamefont {R.~B.}\ \bibnamefont
  {Laughlin}},\ }\href {\doibase 10.1103/PhysRevB.23.5632} {\bibfield
  {journal} {\bibinfo  {journal} {Phys. Rev. B}\ }\textbf {\bibinfo {volume}
  {23}},\ \bibinfo {pages} {5632} (\bibinfo {year} {1981})}\BibitemShut
  {NoStop}%
\bibitem [{\citenamefont {Brouwer}(1998)}]{Brouwer1998}%
  \BibitemOpen
  \bibfield  {author} {\bibinfo {author} {\bibfnamefont {P.~W.}\ \bibnamefont
  {Brouwer}},\ }\href {\doibase 10.1103/PhysRevB.58.R10135} {\bibfield
  {journal} {\bibinfo  {journal} {Phys. Rev. B}\ }\textbf {\bibinfo {volume}
  {58}},\ \bibinfo {pages} {R10135} (\bibinfo {year} {1998})}\BibitemShut
  {NoStop}%
\bibitem [{\citenamefont {Nazarov}\ and\ \citenamefont
  {Blanter}(2009)}]{Nazarov}%
  \BibitemOpen
  \bibfield  {author} {\bibinfo {author} {\bibfnamefont {Y.~V.}\ \bibnamefont
  {Nazarov}}\ and\ \bibinfo {author} {\bibfnamefont {Y.~M.}\ \bibnamefont
  {Blanter}},\ }\href@noop {} {\emph {\bibinfo {title} {Quantum transport:
  introduction to nanoscience}}}\ (\bibinfo  {publisher} {Cambridge University
  Press},\ \bibinfo {year} {2009})\BibitemShut {NoStop}%
\bibitem [{\citenamefont {Altland}\ and\ \citenamefont
  {Zirnbauer}(1997)}]{Altland97}%
  \BibitemOpen
  \bibfield  {author} {\bibinfo {author} {\bibfnamefont {A.}~\bibnamefont
  {Altland}}\ and\ \bibinfo {author} {\bibfnamefont {M.~R.}\ \bibnamefont
  {Zirnbauer}},\ }\href {\doibase 10.1103/PhysRevB.55.1142} {\bibfield
  {journal} {\bibinfo  {journal} {Phys. Rev. B}\ }\textbf {\bibinfo {volume}
  {55}},\ \bibinfo {pages} {1142} (\bibinfo {year} {1997})}\BibitemShut
  {NoStop}%
\bibitem [{\citenamefont {Beenakker}(2015)}]{Beenakker2015}%
  \BibitemOpen
  \bibfield  {author} {\bibinfo {author} {\bibfnamefont {C.~W.~J.}\
  \bibnamefont {Beenakker}},\ }\href {\doibase 10.1103/RevModPhys.87.1037}
  {\bibfield  {journal} {\bibinfo  {journal} {Rev. Mod. Phys.}\ }\textbf
  {\bibinfo {volume} {87}},\ \bibinfo {pages} {1037} (\bibinfo {year}
  {2015})}\BibitemShut {NoStop}%
\bibitem [{\citenamefont {Zak}(1989)}]{Zak1989}%
  \BibitemOpen
  \bibfield  {author} {\bibinfo {author} {\bibfnamefont {J.}~\bibnamefont
  {Zak}},\ }\href {\doibase 10.1103/PhysRevLett.62.2747} {\bibfield  {journal}
  {\bibinfo  {journal} {Phys. Rev. Lett.}\ }\textbf {\bibinfo {volume} {62}},\
  \bibinfo {pages} {2747} (\bibinfo {year} {1989})}\BibitemShut {NoStop}%
\bibitem [{\citenamefont {Ryu}\ and\ \citenamefont {Hatsugai}(2002)}]{Ryu2002}%
  \BibitemOpen
  \bibfield  {author} {\bibinfo {author} {\bibfnamefont {S.}~\bibnamefont
  {Ryu}}\ and\ \bibinfo {author} {\bibfnamefont {Y.}~\bibnamefont {Hatsugai}},\
  }\href {\doibase 10.1103/PhysRevLett.89.077002} {\bibfield  {journal}
  {\bibinfo  {journal} {Phys. Rev. Lett.}\ }\textbf {\bibinfo {volume} {89}},\
  \bibinfo {pages} {077002} (\bibinfo {year} {2002})}\BibitemShut {NoStop}%
\bibitem [{\citenamefont {Qi}\ \emph {et~al.}(2010)\citenamefont {Qi},
  \citenamefont {Hughes},\ and\ \citenamefont {Zhang}}]{Qi2010}%
  \BibitemOpen
  \bibfield  {author} {\bibinfo {author} {\bibfnamefont {X.-L.}\ \bibnamefont
  {Qi}}, \bibinfo {author} {\bibfnamefont {T.~L.}\ \bibnamefont {Hughes}}, \
  and\ \bibinfo {author} {\bibfnamefont {S.-C.}\ \bibnamefont {Zhang}},\ }\href
  {\doibase 10.1103/PhysRevB.81.134508} {\bibfield  {journal} {\bibinfo
  {journal} {Phys. Rev. B}\ }\textbf {\bibinfo {volume} {81}},\ \bibinfo
  {pages} {134508} (\bibinfo {year} {2010})}\BibitemShut {NoStop}%
\bibitem [{\citenamefont {Fu}\ and\ \citenamefont {Kane}(2006)}]{Fu2006}%
  \BibitemOpen
  \bibfield  {author} {\bibinfo {author} {\bibfnamefont {L.}~\bibnamefont
  {Fu}}\ and\ \bibinfo {author} {\bibfnamefont {C.~L.}\ \bibnamefont {Kane}},\
  }\href {\doibase 10.1103/PhysRevB.74.195312} {\bibfield  {journal} {\bibinfo
  {journal} {Phys. Rev. B}\ }\textbf {\bibinfo {volume} {74}},\ \bibinfo
  {pages} {195312} (\bibinfo {year} {2006})}\BibitemShut {NoStop}%
\end{thebibliography}

\end{document}